\documentclass[11pt]{article}
\usepackage{}
\usepackage{amssymb}
\usepackage{amsfonts}
\usepackage{mathrsfs,psfrag,eepic,epsfig}
\usepackage{graphicx}
\usepackage{amsmath}
\usepackage{color}
\usepackage{amsthm}
\usepackage{txfonts}
\usepackage{pifont,bm}
\usepackage{tikz}
\usepackage{epstopdf}
\usepackage{enumerate}
\usepackage{lineno,hyperref}
\modulolinenumbers[5]

\newtheorem{thm}{Theorem}[section]

\theoremstyle{definition}

\makeatletter
\def\@biblabel#1{[#1]}
\makeatother

\makeatletter \@addtoreset{equation}{section}

\begin{document}
%\begin{CJK*}{GBK}{song}

\begin{titlepage}
\title{\bf{Long-time asymptotics in the modified  Landau-Lifshitz equation with nonzero boundary conditions
\footnote{%Project supported by the Fundamental Research Fund  for Talents Cultivation Project of the China University of Mining and Technology under grant number YC150003.\protect\\
%\hspace*{3ex} $^{*}$
Corresponding author.\protect\\
\hspace*{3ex} \emph{E-mail addresses}: wqpeng@cumt.edu.cn (W.Q. Peng) and sftian@cumt.edu.cn,
shoufu2006@126.com (S. F. Tian)}
}}
\author{Wei-Qi Peng and Shou-Fu Tian$^{*}$\\
%%%%%%%%%%%%%%%%%%%%%%%%%%%%%%%%%%%%%%%%%%%%%%%%%%%%%%%%%%%%%%%%%%%%%%%%%%%%%%%%%%%%%%%%%
%%%%%              ÒÔÏÂÁ½ÐÐΪ×÷Õßµ¥Î»
%%%%%%%%%%%%%%%%%%%%%%%%%%%%%%%%%%%%%%%%%%%%%%%%%%%%%%%%%%%%%%%%%%%%%%%%%%%%%%%%%%%%%%%%%
\small \emph{School of Mathematics and Institute of Mathematical Physics, China University of Mining } \\
\small \emph{and Technology, Xuzhou 221116, People's Republic of China}\\
\date{}}

\thispagestyle{empty}
\end{titlepage}
\maketitle

\vspace{-0.5cm}
\begin{center}
\rule{15cm}{1pt}\vspace{0.3cm}

\parbox{15cm}{\small
{\bf Abstract}\\
\hspace{0.5cm}  In this work, we consider the long-time asymptotics of the modified  Landau-Lifshitz equation with nonzero boundary conditions (NZBCs) at infinity. The critical technique is the  deformations of the corresponding matrix Riemann-Hilbert problem via the  nonlinear steepest descent
method, as well as we employ the $g$-function mechanism to eliminate the exponential growths of the jump
matrices. The results indicate that the solution of the modified  Landau-Lifshitz equation with nonzero boundary conditions admits two different asymptotic behavior corresponding to  two types of
regions in the $xt$-plane. They are called the plane wave region with $x<(\beta-4\sqrt{2}q_{0})t, x>(\beta+4\sqrt{2}q_{0})t$, and the modulated elliptic wave region with $(\beta-4\sqrt{2}q_{0})t<x< (\beta+4\sqrt{2}q_{0})t$, respectively.}

\vspace{0.5cm}
\parbox{15cm}{\small{

\vspace{0.3cm} \emph{Key words:}  Long-time asymptotics;  Modified  Landau-Lifshitz equation; Riemann-Hilbert problem; Nonlinear steepest descent method.%\\
%\emph{PACS numbers:}  02.30.Ik, 05.45.Yv, 04.20.Jb. }
}}
\end{center}
\vspace{0.3cm} \rule{15cm}{1pt} \vspace{0.2cm}

%\linenumbers ÐкÅ

\tableofcontents

\section{Introduction}
The modified  Landau-Lifshitz equation which can
be employed to depict the dynamic behavior of local magnetization in electromagnetics takes of the form\cite{Yang1,Yang2}
\begin{align}\label{1}
\frac{\partial \textbf{P}}{\partial t}=-\varphi\textbf{P}\times \textbf{H}_{eff}+\frac{\upsilon}{M_{s}}\textbf{P}\times \frac{\partial \textbf{P}}{\partial t}+\tau_{b},
\end{align}
where $\textbf{P}\equiv\textbf{P}(x,t)$ represents the localized magnetization, the Gilbert damping parameter is be expressed by $\upsilon$, and  $\varphi$ means the gyromagnetic ratio. The effective magnetic field $\textbf{H}_{eff}$ can be redefined as
$\textbf{H}_{eff}=(2E/M_{s}^{2})\partial^{2}\textbf{P}/\partial x^{2}+[(H_{K}/M_{s}-4\pi)M_{\ell}+H_{ext}]\textbf{e}_{\ell}$ which contains
the exchange constant $E$, the anisotropy field $H_{K}$, the applied external field $H_{ext}$, and the unit vector $\textbf{e}_{\ell}$  along the
$\ell$ direction. Taking $\textbf{p}=\textbf{P}/M_{s}$ and making $l_{0}=\sqrt{2E/[(H_{K}-4\pi M_{s})M_{s}]}, t_{0}=1/[\varphi(H_{K}-4\pi M_{s})]$ to rescale the coordinates $x, t$ respectively, we can reduce the Eq.\eqref{1} into the following dimensionless form
\begin{align}\label{2}
\frac{\partial \textbf{p}}{\partial t}=-\textbf{p}\times \frac{\partial^{2} \textbf{p}}{\partial x^{2}}+\upsilon\textbf{p}\times\frac{\partial \textbf{p}}{\partial t}+\frac{b_{J}t_{0}}{l_{0}}\frac{\partial \textbf{p}}{\partial x}-\left(p_{\ell}+\frac{H_{ext}}{H_{K}-4\pi M_{s}}\right)\left(\textbf{p}\times\textbf{e}_{\ell}\right).
\end{align}

Obviously, $\textbf{p}\equiv(p_{x}, p_{y}, p_{\ell})=(0, 0, 1)$ constitutes the ground state of the system, and there are two nonlinear excited states. Since the magnetization of the excited state has little deviation from the ground state as the magnetic field being large enough, we can perform a suitable transformation
\begin{align}\label{2.1}
u=p_{x}+ip_{y},\qquad p_{\ell}=\sqrt{1-\mid u\mid^{2}}.
\end{align}
Substitution of Eq.\eqref{2.1} into Eq.\eqref{2} leads to
\begin{align}\label{2.2}
i\frac{\partial u}{\partial t}=p_{\ell}\frac{\partial^{2} u}{\partial x^{2}}-u\frac{\partial^{2} p_{\ell}}{\partial x^{2}}-\upsilon\left(p_{\ell}\frac{\partial u}{\partial t}-u\frac{\partial p}{\partial t}\right)+i\frac{b_{J}t_{0}}{l_{0}}\frac{\partial u}{\partial x}-\left(p_{\ell}+\frac{H_{ext}}{H_{K}-4\pi M_{s}}\right).
\end{align}

As a non-integrable equation, \eqref{2.2}  can reduce to  the following  integrable equation by  considering the undamped case  and the long wavelength approximation\cite{Yang3} and only retaining the nonlinear terms of the order of magnitude of $\mid u\mid^{2}u$
\begin{align}\label{3}
i\frac{\partial u}{\partial t}-\frac{\partial^{2} u}{\partial x^{2}}-\frac{1}{2}\mid u\mid^{2}u+\left(1+\frac{H_{ext}}{H_{K}-4\pi M_{s}}\right) u-i\frac{b_{J}t_{0}}{l_{0}}\frac{\partial u}{\partial x}=0.
\end{align}
Setting $1+\frac{H_{ext}}{H_{K}-4\pi M_{s}}=\alpha$, $\frac{b_{J}t_{0}}{l_{0}}=\beta$ and $u=2q$, we have
\begin{align}\label{3.1}
iq_{t}-q_{xx}-2\mid q\mid^{2}q+\alpha q-i\beta q_{x}=0,
\end{align}
where the subscripts mean partial derivatives. The Eq.\eqref{3.1} is an absolutely integrable model possessing the soliton solutions\cite{Yang4,Yang5,Yang6}, rogue wave solutions, conservation laws, and modulation instability\cite{Yang7}. To analyze  the characteristics of the soliton solution  in the context of spin waves, the general  soliton expression is constructed by Darboux transformation\cite{Yang9}.  The research for Eq.\eqref{3.1} reveals the accumulation of energy plays a crucial role in the generation of magnetic rogue waves\cite{Yang10}. Besides, different kinds of soliton solutions for Eq.\eqref{3.1} under non-zero boundary conditions are expected to be obtained via the Riemann-Hilbert(RH) problem\cite{Yang}.

In the past years, the asymptotic behavior of solutions has been an significant research topic, which has been reported in a large number of literatures\cite{Peng28,Peng29,Peng30,Peng31,Peng32,Peng33}. In 1993, inspired by classical steepest descent method and earlier work of Its\cite{Peng33}, Deift and Zhou advocated the nonlinear steepest descent method to discuss the long-time asymptotic behavior for the mKdV equation with a oscillatory Riemann-Hilbert problem \cite{Peng34}. Later on,  this method had a further development in Refs.\cite{Peng35,Peng36,Peng37}. Nowadays, the nonlinear steepest descent method have been used to asymptotically analyze lots of integrable equations, such as the non-focusing NLS equation\cite{Peng38}, the KdV equation\cite{Peng39}, the sine-Gordon equation\cite{Peng40}, the Cammasa-Holm equation\cite{Peng41}, the Hirota equation\cite{Peng43,Peng44}, the Kundu-Eckhaus equation\cite{Wang-ke}, coupled NLS equations\cite{Geng} et al. As well as, the long-time asymptotics of the solution  with shock problem\cite{Peng45}, with the time-periodic boundary condition\cite{Peng46,Peng47}, and with the steplike initial data\cite{Peng48,Peng49,Peng50} have been studied. Moreover, as a important development of Riemann-Hilbert problem,  Fokas method was raised to solve the boundary value problems for integrable nonlinear evolution equations\cite{Peng51,Peng52,Peng53,Tian-jde,Ma-jgp}.

Recent years, the researches about nonzero boundary conditions at infinity  have
already been become a hot topic. Biondini and his cooperators have studied the soliton solutions and the long-time asymptotics for the focusing NLS equation with nonzero boundary conditions in \cite{Liunan5} and \cite{Liunan7}, respectively. Very recently,
Tian and his cooperators have reported the soliton solutions for modified Landau-Lifshitz equation \cite{Yang},  focusing Kundu-Eckhaus equation \cite{Yang-ke}, the NLSLab equation \cite{Mao-nc} and  higher-order dispersive NLS quation \cite{Li-nc} with nonzero boundary conditions.
 After that, long-time asymptotics of the focusing Kundu-Eckhaus
equation with nonzero boundary conditions were studied in \cite{Wang}, and long-time dynamics of the Gerdjikov-Ivanov type derivative nonlinear Schr\"{o}dinger equation with nonzero
boundary conditions were studied in \cite{Liunan}. Besides, for the multi-component systems, through employing the theory of inverse scattering transform,  the focusing Manakov system and the three-component defocusing NLS equation with nonzero boundary conditions at infinity has been discussed in \cite{Liunan27} and \cite{Liunan6}, respectively.

In this work, motivated by the long-time asymptotic analysis presented in \cite{Liunan7},  we consider the long-time asymptotics of Eq.\eqref{3.1} with the following nonzero
asymptotic boundary conditions at infinity
\begin{align}\label{4}
q(x,t)\sim q_{\pm}e^{-i\beta x},\qquad x\rightarrow\pm \infty,
\end{align}
where $q_{\pm}$ is independent of $x, t$ and $\mid q_{\pm}\mid=q_{0}>0$. To the best knowledge of the authors, much research work has been done for the Eq.\eqref{3.1}. However, the long-time asymptotics under the nonzero boundary conditions has never been reported up to now.

\textbf{The major results of this work is summarized in what follows:}

\begin{thm}\label{thm-1.1}
For the plane wave sector I, i.e. $\xi>\xi_{2}=\beta+4\sqrt{2}q_{0}$, as $t\rightarrow\infty$,  the long-time asymptotics of $q(x,t)$ is given by
\begin{align}\label{4.1}
q(x,t)=q_{-}e^{i[2g(\infty)-\beta x+(\alpha-2q_{0}^{2})t]}+\mathcal{O}(t^{-\frac{1}{2}}),
\end{align}
where $g(\infty)$ is shown in Eq.\eqref{55}.
\end{thm}
\begin{thm}\label{thm-1.2}
For the modulated elliptic wave sector, i.e. $\beta-4\sqrt{2}q_{0}=\xi_{1}<\xi<\xi_{2}=\beta+4\sqrt{2}q_{0}$, as $t\rightarrow\infty$,  the long-time asymptotics of $q(x,t)$ is given by
\begin{gather}
q(x,t)=\frac{q_{0}(\chi_{2}+q_{0})}{q_{-}^{\ast}}\frac{\Theta\left(\frac{\Lambda t+\vartheta+i\ln(\frac{q_{-}^{\ast}}{iq_{0}})}{2\pi}-V(\infty)+C\right)\Theta(V(\infty)+C)}
{\Theta\left(\frac{\Lambda t+\vartheta+i\ln(\frac{q_{-}^{\ast}}{iq_{0}})}{2\pi}+V(\infty)+C\right)\Theta(-V(\infty)+C)}\notag\\
e^{i[2\tilde{g}(\infty)-\beta x+(\alpha-2q_{0}^{2}-2G(\infty))t]}+\mathcal{O}(t^{-\frac{1}{2}}),\label{4.2}
\end{gather}
where $V(\infty)=\int_{iq_{0}-\frac{\beta}{2}}^{\infty}d\vartheta$, and $\chi_{2}, \Lambda, G(\infty), \vartheta, \tilde{g}(\infty), C$ are given by Eqs. \eqref{84}, \eqref{96}, \eqref{99}, \eqref{103a}, \eqref{103}, \eqref{117a}.
\end{thm}
\begin{thm}\label{thm-1.3}
For the plane wave sector II, i.e. $\xi<\xi_{1}=\beta-4\sqrt{2}q_{0}$, as $t\rightarrow\infty$,  the long-time asymptotics of $q(x,t)$ is given by
\begin{align}\label{4.3}
q(x,t)=q_{+}e^{i[2g(\infty)-\beta x+(\alpha-2q_{0}^{2})t]}+\mathcal{O}(t^{-\frac{1}{2}}),
\end{align}
where $g(\infty)$ is shown in Eq.\eqref{55b}.
\end{thm}

\textbf{Organization of this work:}  In Section 2, in terms of the inverse scattering transform, we derive the basic Riemann-Hilbert problem for
the Eq.\eqref{3.1} with nonzero boundary conditions in Eq.\eqref{4}. In Section 3,  the long-time asymptotics of the solution for Eq.\eqref{3.1} in the plane wave  region I is  computed. In Section 4,  we analyse the  asymptotic behavior in modulated elliptic wave region.  Lastly, by similar procedure, we present the  long-time asymptotics of the solution  in the plane wave  region II.
\section{Reconstructing the basic Riemann-Hilbert problem}

\subsection{Direct scattering problem with NZBCs}
 Eq.\eqref{3.1} admits  the following Lax pair
\begin{align}\label{5}
\phi_{x}=U\phi,\qquad \phi_{t}=V\phi,
\end{align}
with
\begin{gather}
U=ik\sigma_{3}+Q,\notag\\
V=\left[2ik^{2}+i\beta k+\frac{1}{2}\left(i\alpha-2i\mid
q\mid^{2}\right)+iQ_{x}\right]\sigma_{3}+2kQ+\beta Q, \label{6}
\end{gather}
and
\begin{align}\label{7}
Q=\left(\begin{array}{cc}
    0  &  q\\
    -q^{\ast} &  0\\
\end{array}\right),\qquad \sigma_{3}=\left(\begin{array}{cc}
    1  &  0\\
    0 &  -1\\
\end{array}\right),
\end{align}
where $k$ is an spectrum parameter, the superscript $\ast$ represents the complex conjugate,
 and the function $\phi$ is a $2\times 2$ matrix.

Thought the gauge transformation $\phi(x,t)=e^{i[-\beta x+(\alpha-2q_{0}^{2})t]\sigma_{3}/2}\psi(x,t)$, the Lax pair \eqref{5} can turn into
\begin{align}\label{8}
\psi_{x}=X\psi,\qquad \psi_{t}=T\psi,
\end{align}
where
\begin{gather}
X=i(k+\frac{\beta}{2})\sigma_{3}+Q_{1},\qquad Q_{1}=\left(\begin{array}{cc}
    0  &  qe^{-i[-\beta x+(\alpha-2q_{0}^{2})t]}\\
    q^{\ast}e^{i[-\beta x+(\alpha-2q_{0}^{2})t]} &  0\\
\end{array}\right),\notag\\
T=\left[2ik^{2}+i\beta k-i\mid
q\mid^{2}+iq_{0}^{2}\right]\sigma_{3}-iQ_{2}+2kQ_{1}+\beta Q_{1},\notag\\
Q_{2}=\left(\begin{array}{cc}
    0  &  q_{x}e^{-i[-\beta x+(\alpha-2q_{0}^{2})t]}\\
    q^{\ast}_{x}e^{i[-\beta x+(\alpha-2q_{0}^{2})t]} &  0\\
\end{array}\right).\label{9}
\end{gather}
For $x\rightarrow\pm\infty$, defining $Q_{\pm}=\lim\limits_{x\rightarrow \pm\infty}Q$, and using the nonzero asymptotic boundary condition \eqref{4}, the Lax pair in Eq.\eqref{8} becomes
\begin{align}\label{10}
\psi_{\pm x}=X_{\pm}\psi_{\pm},\qquad \psi_{\pm t}=T_{\pm}\psi_{\pm},
\end{align}
where
\begin{gather}
X_{\pm}=i(k+\frac{\beta}{2})\sigma_{3}+Q_{\pm},\quad T_{\pm}=2kX_{\pm}, \notag\\ Q_{\pm}=\left(\begin{array}{cc}
    0  &  q_{\pm}\\
    -q^{\ast}_{\pm} &  0\\
\end{array}\right).
\label{11}
\end{gather}
The eigenvalues of the matrix $X_{\pm}$ are $\pm i\lambda$, where $\lambda$ satisfies
\begin{align}\label{12}
\lambda=\sqrt{(k+\frac{\beta}{2})^{2}+q_{0}^{2}}.
\end{align}
The branch cut of $\lambda$
is $\eta=\eta_{+}\cup \eta_{-}$ with $\eta_{+}=[A,iq_{0}-\frac{\beta}{2}]$ and $\eta_{-}=[-iq_{0}-\frac{\beta}{2},A], A=-\frac{\beta}{2}$, of which the branch cut $\eta$ is oriented upward.(see Fig. 1)\\

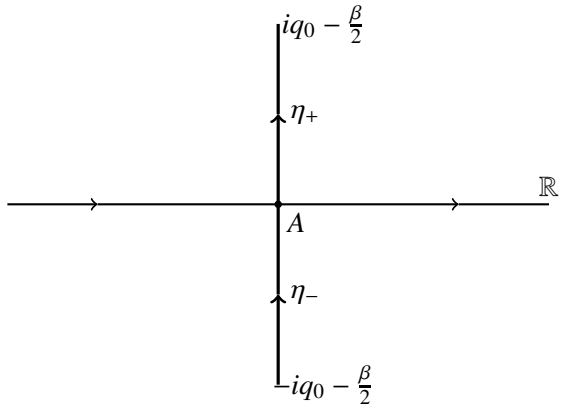
\centerline{\begin{tikzpicture}[scale=1.2]
\draw[->][thick](-3,0)--(-2,0);
\draw[-][thick](-2.0,0)--(0,0);
\draw[fill] (0.2,-0.2) node{$A$};
\draw[->][thick](0,0)--(2,0);
\draw[-][thick](2,0)--(3,0);
\draw[fill] (0,0) circle [radius=0.035];
\draw [->,very thick] (0,0)--(0,1);
\draw [-,very thick] (0,1)--(0,2);
\draw [->,very thick] (0,-2)--(0,-1);
\draw [-,very thick] (0,-1)--(0,0);
\draw[fill] (3,0.2) node{$\mathbb{R}$};
\draw[fill] (0.5,2) node{$iq_{0}-\frac{\beta}{2}$};
\draw[fill] (0.5,-2) node{$-iq_{0}-\frac{\beta}{2}$};
\draw[fill] (0.3,1) node{$\eta_{+}$};
\draw[fill] (0.3,-1) node{$\eta_{-}$};
\end{tikzpicture}}
\centerline{\noindent {\small \textbf{Figure 1.} (Color online) The contour $\Sigma=\mathbb{R}\cup\eta$ of the basic RH problem.}}

The solution of the asymptotic spectral problem \eqref{10} can be derived into
\begin{align}\label{13}
\psi_{\pm}=Y_{\pm}e^{if(x,t,k)\sigma_{3}},
\end{align}
where
\begin{align}\label{14}
f=\lambda\left[x+2kt\right],\quad Y_{\pm}=\left(\begin{array}{cc}
    1  &  \frac{i(\lambda-k-\frac{\beta}{2})}{q^{\ast}_{\pm}}\\
    \frac{i(\lambda-k-\frac{\beta}{2})}{q_{\pm}} &  1\\
\end{array}\right).
\end{align}

Next, we suppose  that $\Psi_{\pm}(x,t,k)$ are both the solutions of the Lax pair in Eq.\eqref{8}, and they meet the asymptotic conditions $\Psi_{\pm}(x,t,k)=\psi_{\pm}(x,t,k)+o(1)$ as $x\rightarrow\infty$. Naturally, $\Psi_{\pm}(x,t,k)$ keep boundedness in $k\in\Sigma=\mathbb{R}\cup\eta$ as $x\rightarrow\pm\infty$, where contour $\Sigma$ is shown in
Fig. 1. Further, by making transformation
\begin{align}\label{15}
\mu_{\pm}(x,t,k)=\Psi_{\pm}(x,t,k)e^{-if\sigma_{3}},
\end{align}
we then get
\begin{align}\label{16}
\mu_{\pm}(x,t,k)=Y_{\pm}+o(1),\qquad x\rightarrow\pm\infty.
\end{align}
Then it is easily to find $\mu_{\pm}$ subject to the following  Lax pair
\begin{gather}
\left(Y^{-1}_{\pm}\mu_{\pm}\right)_{x}+i\lambda[Y^{-1}_{\pm}\mu_{\pm},\sigma_{3}]
=Y^{-1}_{\pm}(Q-Q_{\pm})\mu_{\pm},\notag\\
\left(Y^{-1}_{\pm}\mu_{\pm}\right)_{t}+2i\lambda k[Y^{-1}_{\pm}\mu_{\pm},\sigma_{3}]
=Y^{-1}_{\pm}(T-T_{\pm})\mu_{\pm}.\label{17}
\end{gather}
Eq.\eqref{17} can be written as full derivative form
\begin{align}\label{18}
d\left(e^{if\hat{\sigma}_{3}}Y^{-1}_{\pm}\mu_{\pm}\right)
=e^{if\hat{\sigma}_{3}}\left[Y^{-1}_{\pm}\left(\left(Q-Q_{\pm}\right)dx+\left(T-T_{\pm}\right)dt\right)\mu_{\pm}\right],
\end{align}
which arrives two Volterra integral equations
\begin{align}\label{19}
\mu_{-}(x,t,k)=Y_{-}+\int_{-\infty}^{x}Y_{-}e^{i\lambda(x-y)\hat{\sigma}_{3}}
\left[Y^{-1}_{-}(Q-Q_{-})\mu_{-}(y,t,k)\right]dy,\notag\\
\mu_{+}(x,t,k)=Y_{+}-\int_{x}^{+\infty}Y_{+}e^{i\lambda(x-y)\hat{\sigma}_{3}}
\left[Y^{-1}_{+}(Q-Q_{+})\mu_{+}(y,t,k)\right]dy.
\end{align}
We first suppose $u(x,t)-q_{\pm}\in L^{1}(\mathbb{R}^{\pm})$ and define $\mu_{\pm}=[\mu_{\pm 1},\mu_{\pm 2}]$. The first column of $\mu_{-}$ involves the exponential function $e^{-i\lambda(x-y)}$, which demonstrates that the
first column of $\mu_{-}$ is analytical on  $\mathbb{C}_{-}\setminus \eta_{-}$,  where $\mathbb{C}_{-}=\{k: \mbox{Im}(k)<0\}$. The same argument suggests that the
second column of $\mu_{-}$ is analytical on  $\mathbb{C}_{+}\setminus \eta_{+}$,  where $\mathbb{C}_{+}=\{k: \mbox{Im}(k)>0\}$. Ultimately, we conclude that $\mu_{-2}$ and $\mu_{+1}$ can be analytically continued to $\mathbb{C}_{+}\setminus \eta_{+}$, while $\mu_{-1}$  and $\mu_{+2}$ can be analytically continued to  $\mathbb{C}_{-}\setminus \eta_{-}$.

According to the Abel's theorem, one has
\begin{align}\label{20}
\det\Psi_{\pm}(x,t,k)=\det Y_{\pm}=\frac{2\lambda}{\lambda+ k+\frac{\beta}{2}}\triangleq \Delta(k),
\end{align}
which keeps nonzero and non-singular for $\Sigma_{0}=\Sigma\setminus\{\pm iq_{0}-\frac{\beta}{2}\}$. Since $\Psi_{\pm}(x,t,k)$ satisfy the Lax pair \eqref{8} for $k\in \Sigma_{0}$, the scattering matrix $s(k)$ is defined as
\begin{align}\label{21}
\Psi_{-}(x,t,k)=\Psi_{+}(x,t,k)s(k),\qquad k\in \Sigma_{0}.
\end{align}
We have the symmetry
\begin{align}\label{22}
\Psi^{\ast}_{\pm}(k^{\ast})=-\sigma_{0}\Psi_{\pm}(k)\sigma_{0},\qquad s^{\ast}(k^{\ast})=-\sigma_{0}s(k)\sigma_{0},\qquad k\in \Sigma_{0}.
\end{align}
where $\sigma_{0}=\left(\begin{array}{cc}
   0  &  1\\
    -1 &  0\\
\end{array}\right)$. Thus, the scattering matrix $s(k)$ can be expressed as
\begin{align}\label{23}
s(k)=\left(\begin{array}{cc}
   a(k)  &  -b^{\ast}(k)\\
    b(k) &  a^{\ast}(k)\\
\end{array}\right),\qquad a(k)a^{\ast}(k)+b(k)b^{\ast}(k)=1,
\end{align}
where $a^{\ast}(k)=a^{\ast}(k^{\ast}), b^{\ast}(k)=b^{\ast}(k^{\ast})$ means the Schwartz conjugates. Then we obtain
\begin{align}\label{24}
\Psi_{-1}(x,t,k)=a(k)\Psi_{+1}(x,t,k)+b(k)\Psi_{+2}(x,t,k),\notag\\
\Psi_{-2}(x,t,k)=a^{\ast}(k)\Psi_{+2}(x,t,k)-b^{\ast}(k)\Psi_{+1}(x,t,k).
\end{align}
Considering jumps of the eigenfunctions and scattering data across the branch cut and  taking $\eta$ to be oriented upwards, we define
\begin{align}\label{25}
\mu_{-1}^{-}(k)=\lim\limits_{\varepsilon\rightarrow 0^{+}}\mu_{-1}(k+\varepsilon)=\mu_{-1}(k),\notag\\
\mu_{+2}^{-}(k)=\lim\limits_{\varepsilon\rightarrow 0^{+}}\mu_{+2}(k+\varepsilon)=\mu_{+2}(k), \quad k\in \eta_{-}.
\end{align}
After a direct calculation, we find  the columns of fundamental
solutions $\Psi_{\pm}$ have the following jump conditions across the
branch cut $\eta$, given by
\begin{align}\label{26}
\Psi_{+1}^{+}(x,t,k)=\frac{\lambda+ k+\frac{\beta}{2}}{iq_{+}}\Psi_{+2}(x,t,k),\notag\\
\Psi_{-2}^{+}(x,t,k)=\frac{\lambda+ k+\frac{\beta}{2}}{iq^{\ast}_{-}}\Psi_{-1}(x,t,k),\notag\\
\Psi_{-1}^{+}(x,t,k)=\frac{\lambda+ k+\frac{\beta}{2}}{iq_{-}}\Psi_{-2}(x,t,k),\notag\\
\Psi_{+2}^{+}(x,t,k)=\frac{\lambda+ k+\frac{\beta}{2}}{iq^{\ast}_{+}}\Psi_{+1}(x,t,k),
\end{align}
and the scattering data $a$ yields the following condition across $\eta_{+}$:
\begin{align}\label{27}
(a^{\ast})^{+}(k)=\frac{q_{-}}{q_{+}}a(k),\qquad k\in \eta_{+}.
\end{align}
\subsection{Inverse scattering problem and reconstructing the formula for potential}
First of all, the fundamental matrix-value function can be defined as
\begin{align}\label{28}
m^{(0)}(x, t, k)=\left\{
\begin{array}{lr}
&(\frac{\Psi_{+1}}{a^{\ast}\Delta},\Psi_{-2})e^{-if\sigma_{3}} \qquad k\in\mathbb{C}_{+}\setminus \eta_{+},\\
&(\Psi_{-1},\frac{\Psi_{+2}}{a\Delta})e^{-if\sigma_{3}} \qquad k\in\mathbb{C}_{-}\setminus \eta_{-}.
  \end{array}
\right.
\end{align}
Then the jump condition of the matrix-value function $m^{0}(x, t, k)$ across $\mathbb{R}$ is
\begin{align}\label{29}
m^{(0)}_{+}(x, t, k)=m^{(0)}_{-}(x, t, k)\left(\begin{array}{cc}
   \frac{1}{\Delta}[1+\gamma(k)\gamma^{\ast}(k)]  &  \gamma^{\ast}(k)e^{2if(x,t,k)}\\
    \gamma(k)e^{-2if(x,t,k)} &  \Delta(k)\\
\end{array}\right),\qquad k\in \mathbb{R}.
\end{align}
where $m_{\pm}(x, t, k)$ mean the boundary values of $m(x, t, k)$ in a chosen orientation. The reflection coefficient $\gamma(k)=-\frac{b(k)}{a^{\ast}(k)}$.

Notably, the discontinuities of $\mu_{-1}$ and $\mu_{+2}$ across $\eta_{-}$ and  $\mu_{+1}$ and $\mu_{-2}$ across $\eta_{+}$ will affect the jump of $m$ across the branch cut $\eta$. Therefore, according to \eqref{25}, \eqref{26} and \eqref{27}, we can derive
the jump condition of the matrix-value function $m^{(0)}(x, t, k)$ across $\eta_{+}$, given by
\begin{align}\label{30}
m^{(0)}_{+}(x, t, k)=m^{(0)}_{-}(x, t, k)\left(\begin{array}{cc}
   -\frac{\lambda-k-\beta/2}{iq_{-}}\gamma^{\ast}(k)e^{2if(x,t,k)}  &  \frac{2\lambda}{iq^{\ast}_{-}}\\
     \frac{q^{\ast}_{-}}{2i\lambda}[1+\gamma(k)\gamma^{\ast}(k)]&  -\frac{\lambda+k+\beta/2}{iq^{\ast}_{-}}\gamma(k)e^{-2if(x,t,k)}\\
\end{array}\right),\qquad k\in \eta_{+}.
\end{align}
Analogously, the jump condition of the matrix-value function $m^{(0)}(x, t, k)$ across $\eta_{-}$ is given as follows
\begin{align}\label{31}
m^{(0)}_{+}(x, t, k)=m^{(0)}_{-}(x, t, k)\left(\begin{array}{cc}
   \frac{\lambda+k+\beta/2}{iq_{-}}\gamma^{\ast}(k)e^{2if(x,t,k)}  &   \frac{q_{-}}{2i\lambda}[1+\gamma(k)\gamma^{\ast}(k)]\\
  \frac{2\lambda}{iq_{-}}  &  \frac{\lambda-k-\beta/2}{iq^{\ast}_{-}}\gamma(k)e^{-2if(x,t,k)}\\
\end{array}\right),\qquad k\in \eta_{-}.
\end{align}
Now, to realize that there is no discrete spectrum, we suppose that $a\neq 0$ for all $k\in \mathbb{C}_{-}\cup \Sigma$. Then, the matrix-value function $m^{(0)}(x, t, k)$ is the solution to the following Riemann-Hilbert problem:
\begin{align}\label{32}
\left\{
\begin{array}{lr}
m^{(0)}(x, t, k)\ \mbox{is analytic in} \ \mathbb{C}\setminus\Sigma,\\
m^{(0)}_{+}(x, t, k)=m^{(0)}_{-}(x, t, k)J^{(0)}(x, t, k), \qquad k\in\Sigma,\\
m^{(0)}(x, t, k)\rightarrow I,\qquad k\rightarrow \infty,
  \end{array}
\right.
\end{align}
of which the jump matrix $J^{(0)}(x, t, k)=\{J^{(0)}_{i}(x, t, k)\}_{i=1}^{3}$ read (see Figure 1)
\begin{gather}
J_{1}^{(0)}=\left(\begin{array}{cc}
   \frac{1}{\Delta}[1+\gamma(k)\gamma^{\ast}(k)]  &  \gamma^{\ast}(k)e^{2i\theta t}\\
    \gamma(k)e^{-2i\theta t} &  \Delta(k)\\
\end{array}\right),\notag\\
J_{2}^{(0)}=\left(\begin{array}{cc}
   -\frac{\lambda-k-\beta/2}{iq_{-}}\gamma^{\ast}(k)e^{2i\theta t}  &  \frac{2\lambda}{iq^{\ast}_{-}}\\
     \frac{q^{\ast}_{-}}{2i\lambda}[1+\gamma(k)\gamma^{\ast}(k)]&  -\frac{\lambda+k+\beta/2}{iq^{\ast}_{-}}\gamma(k)e^{-2i\theta t}\\
\end{array}\right),\notag\\
J_{3}^{(0)}=\left(\begin{array}{cc}
   \frac{\lambda+k+\beta/2}{iq_{-}}\gamma^{\ast}(k)e^{2i\theta t}  &   \frac{q_{-}}{2i\lambda}[1+\gamma(k)\gamma^{\ast}(k)]\\
  \frac{2\lambda}{iq_{-}}  &  \frac{\lambda-k-\beta/2}{iq^{\ast}_{-}}\gamma(k)e^{-2i\theta t}\\
\end{array}\right),\notag
\end{gather}
where $\theta=\lambda(\xi+2k), \xi=\frac{x}{t}$.

In order to reconstruct the formula for potential $q(x, t)$, we need to
expand the $M^{(0)}(x, t, k)$ at large $k$ as
\begin{align}\label{33}
m^{(0)}(x, t, k)=I+\frac{m_{1}^{(0)}(x, t)}{k}+\frac{m_{2}^{(0)}(x, t)}{k^{2}}+\mathcal{O}(\frac{1}{k^{3}}),\qquad k\rightarrow \infty.
\end{align}
Combining equations \eqref{8}, \eqref{28}  and \eqref{33}, one can also recover the solution
of the Eq.\eqref{3.1} in the form
\begin{align}\label{34}
q(x,t)=-2i\left(m_{1}^{(0)}(x,t)\right)_{12}e^{i[-\beta x+(\alpha-2q_{0}^{2})t]},
\end{align}
\subsection{The sign structure of $\mbox{Im}(\theta)$}
%First, we need to choose contour deformations,
%which depends crucially on the sign structure of the quantity Im($\theta$). We note that
%\begin{align}\label{35}
%\theta(\xi,k)=\sqrt{(k+\frac{\beta}{2})^{2}+q_{0}^{2}}(\xi+2k).
%\end{align}
%Thus, we get
%\begin{align}\label{36}
%\frac{d\theta(\xi,k)}{dk}=\frac{8k^{2}+2(\xi+3\beta)k+\beta\xi+\beta^{2}+4q_{0}^{2}}{2\lambda},
%\end{align}
%which implies that $\theta$ has two stationary points
%\begin{align}\label{37}
%k_{\pm}=\frac{1}{8}(-3\beta-\xi\pm\sqrt{\xi^{2}-2\beta\xi+\beta^{2}-32q_{0}^{2}}),
%\end{align}
%Letting $k= k_{1}+ik_{2}$, we have
%\begin{align}\label{38}
%\theta^{2}=\left(k_{1}^{2}-k_{2}^{2}+k_{1}\beta+\frac{\beta}{4} +q_{0}^{2}+(2k_{1}k_{2}+k_{2}\beta)i\right)
%\left(4k_{1}^{2}-4k_{2}^{2}+4k_{1}\xi+\xi^{2}+(8k_{1}k_{2}+4k_{2}\xi)i\right),
%\end{align}
%Hence, according to
%\begin{align}\label{39}
%\{\mbox{Im}(\theta)=0\}\equiv \{\mbox{Im}(\theta^{2})=0\}\cap \{\mbox{Re}(\theta^{2})\geq 0\},
%\end{align}
%we can find the curves of the sets $\mbox{Im}(\theta)=0$. In fact, the sign structure of Im($\theta$) in
%the complex $k$-plane is shown in Fig. 2.

Before we start the contour deformations,
we need to discuss the  sign structure of the quantity $\mbox{Im}(\theta)$. Let $\lambda=\lambda_{1}+i\lambda_{2}, k=k_{1}+ik_{2}$, we can get
\begin{align}\label{35}
\mbox{Im}(\theta)=\lambda_{2}(\xi+2k_{1})+2\lambda_{1}k_{2}.
\end{align}
At the case of $\mid k_{2}\mid\gg 1$, since $\lambda\sim k$ as $k\rightarrow\infty$, one obtains
\begin{align}\label{36}
\mbox{Im}(\theta)=(4k_{1}+\xi)k_{2}+\mathcal{O}(\frac{1}{k2}), \quad k_{2}\rightarrow\pm\infty.
\end{align}
Therefore, $4k_{1}+\xi$ determines the sign of $\mbox{Im}(\theta)$ as $k_{2}\rightarrow\pm\infty$. On the other hand, consider $0<k_{2}\ll 1$ and the definition of $\lambda$
\begin{align}\label{37}
\lambda=\mbox{sign}(k_{1})\sqrt{k_{1}^{2}+\beta k_{1}+\frac{\beta^{2}}{4}+q_{0}^{2}}\left[1+\frac{(2k_{1}+\beta)i}{2(k_{1}^{2}+\beta k_{1}+\frac{\beta^{2}}{4}+q_{0}^{2})}k_{2}+\mathcal{O}(k_{2}^{2})\right],
\end{align}
we can easily derive
\begin{align}\label{38}
\mbox{Im}(\theta)=\frac{\mbox{sign}(k_{1})}{2\sqrt{k_{1}^{2}+\beta k_{1}+\frac{\beta^{2}}{4}+q_{0}^{2}}}
\left[8k_{1}^{2}+(6\beta+2\xi)k_{1}+\beta\xi+\beta^{2}+4q_{0}^{2}\right]k_{2}+\mathcal{O}(k_{2}^{3}),\ k_{2}\rightarrow 0_{+}.
\end{align}
Then one has $\mbox{Im}(\theta)=0$ on the real axis if
\begin{align}\label{39}
k_{1}^{\pm}=\frac{1}{8}(-3\beta-\xi\pm\sqrt{\xi^{2}-2\beta\xi+\beta^{2}-32q_{0}^{2}}).
\end{align}
In fact, we can show  the sign structure of Im($\theta$) in
the complex $k$-plane (see Fig. 2).

\noindent\textbf{Remark}: Set $\xi_{1}=\beta-4\sqrt{2}q_{0}, \xi_{2}=\beta+4\sqrt{2}q_{0}$, when $\xi<\xi_{1}, \xi>\xi_{2}$, the function
$\theta$ has two real stationary points corresponding to
plane wave regions. While  $\xi_{1}<\xi<\xi_{2}$, the sector will be  the modulated elliptic
wave regions.\\
\centerline{\begin{tikzpicture}[scale=0.8]
\draw[-][thick] (-7,-2) -- (-6.5,-2);
\draw[-][thick] (-6.5,-2) -- (-6,-2);
\draw[-][thick] (-6,-2) -- (-5.5,-2);
\draw[-][thick] (-5.5,-2) -- (-5,-2);
\draw [thick] (-7,-2) node[below]{$0$} -- (-7,-1.95);
\draw [thick] (-6.5,-2) node[below]{$1$} -- (-6.5,-1.95);
\draw [thick] (-6,-2) node[below]{$2$} -- (-6,-1.95);
\draw [thick] (-5.5,-2) node[below]{$3$} -- (-5.5,-1.95);
\draw [thick] (-5,-2) node[below]{$4$} -- (-5,-1.95);
\draw[-][thick] (-7,-2) -- (-7.5,-2);
\draw[-][thick] (-7.5,-2) -- (-8,-2);
\draw[-][thick] (-8,-2) -- (-8.5,-2);
\draw[-][thick] (-8.5,-2) -- (-9,-2);
\draw [thick] (-7.5,-2) node[below]{$-1$} -- (-7.5,-1.95);
\draw [thick] (-8,-2) node[below]{$-2$} -- (-8,-1.95);
\draw [thick] (-8.5,-2) node[below]{$-3$} -- (-8.5,-1.95);
\draw [thick] (-9,-2) node[below]{$-4$} -- (-9,-1.95);
\draw[-][thick] (-9,0) -- (-5,0);
\draw[-][thick] (-7,2) -- (-6.5,2);
\draw[-][thick] (-6.5,2) -- (-6,2);
\draw[-][thick] (-6,2) -- (-5.5,2);
\draw[-][thick] (-5.5,2) -- (-5,2);
\draw [thick] (-7,2)  -- (-7,1.95);
\draw [thick] (-6.5,2)  -- (-6.5,1.95);
\draw [thick] (-6,2)  -- (-6,1.95);
\draw [thick] (-5.5,2)  -- (-5.5,1.95);
\draw [thick] (-5,2)  -- (-5,1.95);
\draw[-][thick] (-7,2) -- (-7.5,2);
\draw[-][thick] (-7.5,2) -- (-8,2);
\draw[-][thick] (-8,2) -- (-8.5,2);
\draw[-][thick] (-8.5,2) -- (-9,2);
\draw [thick] (-7.5,2)  -- (-7.5,1.95);
\draw [thick] (-8,2)  -- (-8,1.95);
\draw [thick] (-8.5,2)  -- (-8.5,1.95);
\draw [thick] (-9,2)  -- (-9,1.95);
\draw[-][thick] (-9,-2) -- (-9,-1.5);
\draw[-][thick] (-9,-1.5) -- (-9,-1);
\draw[-][thick] (-9,-1) -- (-9,-0.5);
\draw[-][thick] (-9,-0.5) -- (-9,0);
\draw[-][thick] (-9,0) -- (-9,0.5);
\draw[-][thick] (-9,0.5) -- (-9,1);
\draw[-][thick] (-9,1) -- (-9,1.5);
\draw[-][thick] (-9,1.5) -- (-9,2);
\draw [thick] (-9,-2) node[left]{$-4$};
\draw [thick] (-9,-1.5) node[left]{$-3$} -- (-8.95,-1.5);
\draw [thick] (-9,-1) node[left]{$-2$} -- (-8.95,-1);
\draw [thick] (-9,-0.5) node[left]{$-1$} -- (-8.95,-0.5);
\draw [thick] (-9,0) node[left]{$0$} -- (-8.95,0);
\draw [thick] (-9,0.5) node[left]{$1$} -- (-8.95,0.5);
\draw [thick] (-9,1) node[left]{$2$} -- (-8.95,1);
\draw [thick] (-9,1.5) node[left]{$3$} -- (-8.95,1.5);
\draw [thick] (-9,2) node[left]{$4$} -- (-8.95,2);
\draw[-][thick] (-5,-2) -- (-5,-1.5);
\draw[-][thick] (-5,-1.5) -- (-5,-1);
\draw[-][thick] (-5,-1) -- (-5,-0.5);
\draw[-][thick] (-5,-0.5) -- (-5,0);
\draw[-][thick] (-5,0) -- (-5,0.5);
\draw[-][thick] (-5,0.5) -- (-5,1);
\draw[-][thick] (-5,1) -- (-5,1.5);
\draw[-][thick] (-5,1.5) -- (-5,2);
\draw [thick] (-5,-1.5)  -- (-5.05,-1.5);
\draw [thick] (-5,-1)  -- (-5.05,-1);
\draw [thick] (-5,-0.5) -- (-5.05,-0.5);
\draw [thick] (-5,0)  -- (-5.05,0);
\draw [thick] (-5,0.5)  -- (-5.05,0.5);
\draw [thick] (-5,1)  -- (-5.05,1);
\draw [thick] (-5,1.5)  -- (-5.05,1.5);
\draw [thick] (-5,2)  -- (-5.05,2);
\draw[-] [dashed,thick] (-7.5,0.5) -- (-7.5,-0.5);
\draw [-,thick] (-7.75,0) to [out=90,in=-120] (-7.5,0.5);
\draw [-,thick] (-7.75,0) to [out=-90,in=120] (-7.5,-0.5);
\draw [-,thick] (-8,0) to [out=95,in=-85] (-8.375,2);
\draw [-,thick] (-8,0) to [out=-95,in=85] (-8.375,-2);
\draw[fill] (-8.5,0.5) node{$\textcolor[rgb]{1.00,0.00,0.00}{\textbf{--}}$\ $\textcolor[rgb]{1.00,0.00,0.00}{\textbf{--}}$};
\draw[fill] (-8.5,-0.5) node{ $\textcolor[rgb]{1.00,0.00,0.00}{\textbf{++}}$};
\draw[fill] (-6.5,1) node{ $\textcolor[rgb]{1.00,0.00,0.00}{\textbf{++}}$};
\draw[fill] (-6.5,-1) node{$\textcolor[rgb]{1.00,0.00,0.00}{\textbf{--}}$\ $\textcolor[rgb]{1.00,0.00,0.00}{\textbf{--}}$};
\draw[-][thick] (-2,-2) -- (-1.5,-2);
\draw[-][thick] (-1.5,-2) -- (-1,-2);
\draw[-][thick] (-1,-2) -- (-0.5,-2);
\draw[-][thick] (-0.5,-2) -- (0,-2);
\draw [thick] (-2,-2) node[below]{$0$} -- (-2,-1.95);
\draw [thick] (-1.5,-2) node[below]{$1$} -- (-1.5,-1.95);
\draw [thick] (-1,-2) node[below]{$2$} -- (-1,-1.95);
\draw [thick] (-0.5,-2) node[below]{$3$} -- (-0.5,-1.95);
\draw [thick] (0,-2) node[below]{$4$} -- (0,-1.95);
\draw[-][thick] (-2,-2) -- (-2.5,-2);
\draw[-][thick] (-2.5,-2) -- (-3,-2);
\draw[-][thick] (-3,-2) -- (-3.5,-2);
\draw[-][thick] (-3.5,-2) -- (-4,-2);
\draw [thick] (-2.5,-2) node[below]{$-1$} -- (-2.5,-1.95);
\draw [thick] (-3,-2) node[below]{$-2$} -- (-3,-1.95);
\draw [thick] (-3.5,-2) node[below]{$-3$} -- (-3.5,-1.95);
\draw [thick] (-4,-2) node[below]{$-4$} -- (-4,-1.95);
\draw[-][thick] (-4,0) -- (0,0);
\draw[-][thick] (-2,2) -- (-1.5,2);
\draw[-][thick] (-1.5,2) -- (-1,2);
\draw[-][thick] (-1,2) -- (-0.5,2);
\draw[-][thick] (-0.5,2) -- (0,2);
\draw [thick] (-2,2)  -- (-2,1.95);
\draw [thick] (-1.5,2)  -- (-1.5,1.95);
\draw [thick] (-1,2)  -- (-1,1.95);
\draw [thick] (-0.5,2)  -- (-0.5,1.95);
\draw [thick] (0,2)  -- (0,1.95);
\draw[-][thick] (-2,2) -- (-2.5,2);
\draw[-][thick] (-2.5,2) -- (-3,2);
\draw[-][thick] (-3,2) -- (-3.5,2);
\draw[-][thick] (-3.5,2) -- (-4,2);
\draw [thick] (-2.5,2)  -- (-2.5,1.95);
\draw [thick] (-3,2)  -- (-3,1.95);
\draw [thick] (-3.5,2)  -- (-3.5,1.95);
\draw [thick] (-4,2)  -- (-4,1.95);
\draw[-][thick] (-4,-2) -- (-4,-1.5);
\draw[-][thick] (-4,-1.5) -- (-4,-1);
\draw[-][thick] (-4,-1) -- (-4,-0.5);
\draw[-][thick] (-4,-0.5) -- (-4,0);
\draw[-][thick] (-4,0) -- (-4,0.5);
\draw[-][thick] (-4,0.5) -- (-4,1);
\draw[-][thick] (-4,1) -- (-4,1.5);
\draw[-][thick] (-4,1.5) -- (-4,2);
\draw [thick] (-4,-2) node[left]{$-4$};
\draw [thick] (-4,-1.5) node[left]{$-3$} -- (-3.95,-1.5);
\draw [thick] (-4,-1) node[left]{$-2$} -- (-3.95,-1);
\draw [thick] (-4,-0.5) node[left]{$-1$} -- (-3.95,-0.5);
\draw [thick] (-4,0) node[left]{$0$} -- (-3.95,0);
\draw [thick] (-4,0.5) node[left]{$1$} -- (-3.95,0.5);
\draw [thick] (-4,1) node[left]{$2$} -- (-3.95,1);
\draw [thick] (-4,1.5) node[left]{$3$} -- (-3.95,1.5);
\draw [thick] (-4,2) node[left]{$4$} -- (-3.95,2);
\draw[-][thick] (0,-2) -- (0,-1.5);
\draw[-][thick] (0,-1.5) -- (0,-1);
\draw[-][thick] (0,-1) -- (0,-0.5);
\draw[-][thick] (0,-0.5) -- (0,0);
\draw[-][thick] (0,0) -- (0,0.5);
\draw[-][thick] (0,0.5) -- (0,1);
\draw[-][thick] (0,1) -- (0,1.5);
\draw[-][thick] (0,1.5) -- (0,2);
\draw [thick] (0,-1.5)  -- (-0.05,-1.5);
\draw [thick] (0,-1)  -- (-0.05,-1);
\draw [thick] (0,-0.5) -- (-0.05,-0.5);
\draw [thick] (0,0)  -- (-0.05,0);
\draw [thick] (0,0.5)  -- (-0.05,0.5);
\draw [thick] (0,1)  -- (-0.05,1);
\draw [thick] (0,1.5)  -- (-0.05,1.5);
\draw [thick] (0,2)  -- (-0.05,2);
\draw[-] [dashed,thick] (-2.5,0.5) -- (-2.5,-0.5);
\draw [-,thick] (-2.5,0.5) to [out=-135,in=0] (-2.86,0.25);
\draw [-,thick] (-2.5,-0.5) to [out=135,in=0] (-2.86,-0.25);
\draw [-,thick] (-2.86,0.25) to [out=180,in=-80] (-3.325,2);
\draw [-,thick] (-2.86,-0.25) to [out=-180,in=80] (-3.325,-2);
\draw[fill] (-3.6,0.5) node{$\textcolor[rgb]{1.00,0.00,0.00}{\textbf{--}}$\ $\textcolor[rgb]{1.00,0.00,0.00}{\textbf{--}}$};
\draw[fill] (-3.6,-0.5) node{ $\textcolor[rgb]{1.00,0.00,0.00}{\textbf{++}}$};
\draw[fill] (-1.5,1) node{ $\textcolor[rgb]{1.00,0.00,0.00}{\textbf{++}}$};
\draw[fill] (-1.5,-1) node{$\textcolor[rgb]{1.00,0.00,0.00}{\textbf{--}}$\ $\textcolor[rgb]{1.00,0.00,0.00}{\textbf{--}}$};
\draw[-][thick] (3,-2) -- (3.5,-2);
\draw[-][thick] (3.5,-2) -- (4,-2);
\draw[-][thick] (4,-2) -- (4.5,-2);
\draw[-][thick] (4.5,-2) -- (5,-2);
\draw [thick] (3,-2) node[below]{$0$} -- (3,-1.95);
\draw [thick] (3.5,-2) node[below]{$1$} -- (3.5,-1.95);
\draw [thick] (4,-2) node[below]{$2$} -- (4,-1.95);
\draw [thick] (4.5,-2) node[below]{$3$} -- (4.5,-1.95);
\draw [thick] (5,-2) node[below]{$4$} -- (5,-1.95);
\draw[-][thick] (3,-2) -- (2.5,-2);
\draw[-][thick] (2.5,-2) -- (2,-2);
\draw[-][thick] (2,-2) -- (1.5,-2);
\draw[-][thick] (1.5,-2) -- (1,-2);
\draw [thick] (2.5,-2) node[below]{$-1$} -- (2.5,-1.95);
\draw [thick] (2,-2) node[below]{$-2$} -- (2,-1.95);
\draw [thick] (1.5,-2) node[below]{$-3$} -- (1.5,-1.95);
\draw [thick] (1,-2) node[below]{$-4$} -- (1,-1.95);
\draw[-][thick] (1,0) -- (5,0);
\draw[-][thick] (3,2) -- (3.5,2);
\draw[-][thick] (3.5,2) -- (4,2);
\draw[-][thick] (4,2) -- (4.5,2);
\draw[-][thick] (4.5,2) -- (5,2);
\draw [thick] (3,2)  -- (3,1.95);
\draw [thick] (3.5,2)  -- (3.5,1.95);
\draw [thick] (4,2)  -- (4,1.95);
\draw [thick] (4.5,2)  -- (4.5,1.95);
\draw [thick] (5,2)  -- (5,1.95);
\draw[-][thick] (3,2) -- (2.5,2);
\draw[-][thick] (2.5,2) -- (2,2);
\draw[-][thick] (2,2) -- (1.5,2);
\draw[-][thick] (1.5,2) -- (1,2);
\draw [thick] (2.5,2)  -- (2.5,1.95);
\draw [thick] (2,2)  -- (2,1.95);
\draw [thick] (1.5,2)  -- (1.5,1.95);
\draw [thick] (1,2)  -- (1,1.95);
\draw[-][thick] (1,-2) -- (1,-1.5);
\draw[-][thick] (1,-1.5) -- (1,-1);
\draw[-][thick] (1,-1) -- (1,-0.5);
\draw[-][thick] (1,-0.5) -- (1,0);
\draw[-][thick] (1,0) -- (1,0.5);
\draw[-][thick] (1,0.5) -- (1,1);
\draw[-][thick] (1,1) -- (1,1.5);
\draw[-][thick] (1,1.5) -- (1,2);
\draw [thick] (1,-2) node[left]{$-4$};
\draw [thick] (1,-1.5) node[left]{$-3$} -- (1.05,-1.5);
\draw [thick] (1,-1) node[left]{$-2$} -- (1.05,-1);
\draw [thick] (1,-0.5) node[left]{$-1$} -- (1.05,-0.5);
\draw [thick] (1,0) node[left]{$0$} -- (1.05,0);
\draw [thick] (1,0.5) node[left]{$1$} -- (1.05,0.5);
\draw [thick] (1,1) node[left]{$2$} -- (1.05,1);
\draw [thick] (1,1.5) node[left]{$3$} -- (1.05,1.5);
\draw [thick] (1,2) node[left]{$4$} -- (1.05,2);
\draw[-][thick] (5,-2) -- (5,-1.5);
\draw[-][thick] (5,-1.5) -- (5,-1);
\draw[-][thick] (5,-1) -- (5,-0.5);
\draw[-][thick] (5,-0.5) -- (5,0);
\draw[-][thick] (5,0) -- (5,0.5);
\draw[-][thick] (5,0.5) -- (5,1);
\draw[-][thick] (5,1) -- (5,1.5);
\draw[-][thick] (5,1.5) -- (5,2);
\draw [thick] (5,-1.5)  -- (4.95,-1.5);
\draw [thick] (5,-1)  -- (4.95,-1);
\draw [thick] (5,-0.5) -- (4.95,-0.5);
\draw [thick] (5,0)  -- (4.95,0);
\draw [thick] (5,0.5)  -- (4.95,0.5);
\draw [thick] (5,1)  -- (4.95,1);
\draw [thick] (5,1.5)  -- (4.95,1.5);
\draw [thick] (5,2)  -- (4.95,2);
\draw[-] [dashed,thick] (2.5,0.5) -- (2.5,-0.5);
\draw [-,thick] (2.5,0.5) to [out=-45,in=180] (2.86,0.25);
\draw [-,thick] (2.5,-0.5) to [out=135,in=0] (2.86,-0.25);
\draw [-,thick] (2.86,0.25) to [out=0,in=-100] (3.325,2);
\draw [-,thick] (2.86,-0.25) to [out=0,in=100] (3.325,-2);
\draw[fill] (1.4,0.5) node{$\textcolor[rgb]{1.00,0.00,0.00}{\textbf{--}}$\ $\textcolor[rgb]{1.00,0.00,0.00}{\textbf{--}}$};
\draw[fill] (1.4,-0.5) node{ $\textcolor[rgb]{1.00,0.00,0.00}{\textbf{++}}$};
\draw[fill] (4,1) node{ $\textcolor[rgb]{1.00,0.00,0.00}{\textbf{++}}$};
\draw[fill] (4,-1) node{$\textcolor[rgb]{1.00,0.00,0.00}{\textbf{--}}$\ $\textcolor[rgb]{1.00,0.00,0.00}{\textbf{--}}$};
\draw[-][thick] (8,-2) -- (8.5,-2);
\draw[-][thick] (8.5,-2) -- (9,-2);
\draw[-][thick] (9,-2) -- (9.5,-2);
\draw[-][thick] (9.5,-2) -- (10,-2);
\draw [thick] (8,-2) node[below]{$0$} -- (8,-1.95);
\draw [thick] (8.5,-2) node[below]{$1$} -- (8.5,-1.95);
\draw [thick] (9,-2) node[below]{$2$} -- (9,-1.95);
\draw [thick] (9.5,-2) node[below]{$3$} -- (9.5,-1.95);
\draw [thick] (10,-2) node[below]{$4$} -- (10,-1.95);
\draw[-][thick] (8,-2) -- (7.5,-2);
\draw[-][thick] (7.5,-2) -- (7,-2);
\draw[-][thick] (7,-2) -- (6.5,-2);
\draw[-][thick] (6.5,-2) -- (6,-2);
\draw [thick] (7.5,-2) node[below]{$-1$} -- (7.5,-1.95);
\draw [thick] (7,-2) node[below]{$-2$} -- (7,-1.95);
\draw [thick] (6.5,-2) node[below]{$-3$} -- (6.5,-1.95);
\draw [thick] (6,-2) node[below]{$-4$} -- (6,-1.95);
\draw[-][thick] (6,0) -- (10,0);
\draw[-][thick] (8,2) -- (8.5,2);
\draw[-][thick] (8.5,2) -- (9,2);
\draw[-][thick] (9,2) -- (9.5,2);
\draw[-][thick] (9.5,2) -- (10,2);
\draw [thick] (8,2)  -- (8,1.95);
\draw [thick] (8.5,2)  -- (8.5,1.95);
\draw [thick] (9,2)  -- (9,1.95);
\draw [thick] (9.5,2)  -- (9.5,1.95);
\draw [thick] (10,2)  -- (10,1.95);
\draw[-][thick] (8,2) -- (7.5,2);
\draw[-][thick] (7.5,2) -- (7,2);
\draw[-][thick] (7,2) -- (6.5,2);
\draw[-][thick] (6.5,2) -- (6,2);
\draw [thick] (7.5,2)  -- (7.5,1.95);
\draw [thick] (7,2)  -- (7,1.95);
\draw [thick] (6.5,2)  -- (6.5,1.95);
\draw [thick] (6,2)  -- (6,1.95);
\draw[-][thick] (6,-2) -- (6,-1.5);
\draw[-][thick] (6,-1.5) -- (6,-1);
\draw[-][thick] (6,-1) -- (6,-0.5);
\draw[-][thick] (6,-0.5) -- (6,0);
\draw[-][thick] (6,0) -- (6,0.5);
\draw[-][thick] (6,0.5) -- (6,1);
\draw[-][thick] (6,1) -- (6,1.5);
\draw[-][thick] (6,1.5) -- (6,2);
\draw [thick] (6,-2) node[left]{$-4$};
\draw [thick] (6,-1.5) node[left]{$-3$} -- (6.05,-1.5);
\draw [thick] (6,-1) node[left]{$-2$} -- (6.05,-1);
\draw [thick] (6,-0.5) node[left]{$-1$} -- (6.05,-0.5);
\draw [thick] (6,0) node[left]{$0$} -- (6.05,0);
\draw [thick] (6,0.5) node[left]{$1$} -- (6.05,0.5);
\draw [thick] (6,1) node[left]{$2$} -- (6.05,1);
\draw [thick] (6,1.5) node[left]{$3$} -- (6.05,1.5);
\draw [thick] (6,2) node[left]{$4$} -- (6.05,2);
\draw[-][thick] (10,-2) -- (10,-1.5);
\draw[-][thick] (10,-1.5) -- (10,-1);
\draw[-][thick] (10,-1) -- (10,-0.5);
\draw[-][thick] (10,-0.5) -- (10,0);
\draw[-][thick] (10,0) -- (10,0.5);
\draw[-][thick] (10,0.5) -- (10,1);
\draw[-][thick] (10,1) -- (10,1.5);
\draw[-][thick] (10,1.5) -- (10,2);
\draw [thick] (10,-1.5)  -- (9.95,-1.5);
\draw [thick] (10,-1)  -- (9.95,-1);
\draw [thick] (10,-0.5) -- (9.95,-0.5);
\draw [thick] (10,0)  -- (9.95,0);
\draw [thick] (10,0.5)  -- (9.95,0.5);
\draw [thick] (10,1)  -- (9.95,1);
\draw [thick] (10,1.5)  -- (9.95,1.5);
\draw [thick] (10,2)  -- (9.95,2);
\draw[-] [dashed,thick] (7.5,0.5) -- (7.5,-0.5);
\draw [-,thick] (7.75,0) to [out=90,in=-60] (7.5,0.5);
\draw [-,thick] (7.75,0) to [out=-90,in=60] (7.5,-0.5);
\draw [-,thick] (8,0) to [out=85,in=-95] (8.375,2);
\draw [-,thick] (8,0) to [out=-85,in=95] (8.375,-2);
\draw[fill] (6.5,0.5) node{$\textcolor[rgb]{1.00,0.00,0.00}{\textbf{--}}$\ $\textcolor[rgb]{1.00,0.00,0.00}{\textbf{--}}$};
\draw[fill] (6.5,-0.5) node{ $\textcolor[rgb]{1.00,0.00,0.00}{\textbf{++}}$};
\draw[fill] (9,1) node{ $\textcolor[rgb]{1.00,0.00,0.00}{\textbf{++}}$};
\draw[fill] (9,-1) node{$\textcolor[rgb]{1.00,0.00,0.00}{\textbf{--}}$\ $\textcolor[rgb]{1.00,0.00,0.00}{\textbf{--}}$};
\end{tikzpicture}}\\
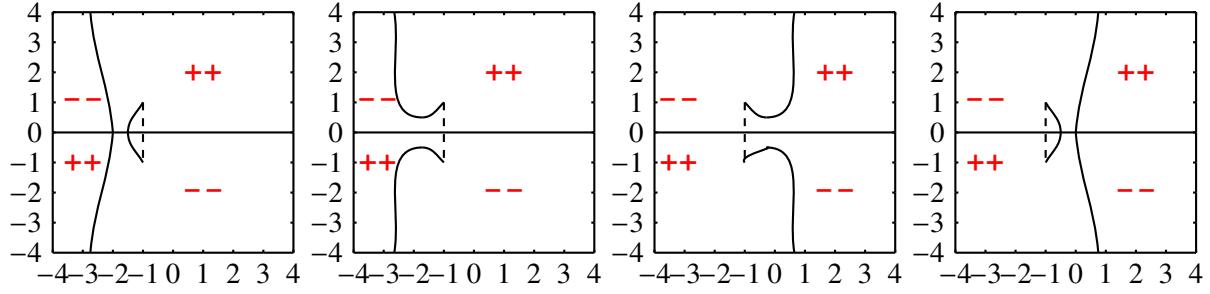
\noindent { \small \textbf{Figure 2.} (Color online) The sign structure of $\mbox{Im}(\theta)=0$ in the complex $k$-plane for various values  $\beta=2, q_{0}=1$ and
$\textbf{(a)}$ $\xi=8$,
$\textbf{(b)}$ $\xi=7.2$,
$\textbf{(c)}$ $\xi=-3.2$,
$\textbf{(d)}$ $\xi=-4$.}\\

\section{Plane wave region I}
For the plane wave region I $\xi>\xi_{2}$, there are two real stationary points for $\theta$. In order to derive long-time asymptotics of solution for the Eq.\eqref{3.1} with boundary conditions \eqref{4}, we carry out similar deformations of the Riemann-Hilbert problem \eqref{32} as that in Refs. \cite{Peng45,Liunan7}.
\subsection{First deformation}
In this subsection, our goal is to realize the deformation between $m^{(0)}$ and $m^{(1)}$. We first decompose the jump matrix $J_{1}^{(0)}, J_{2}^{(0)}, J_{3}^{(0)}$, given in what follows
\begin{align}
\left\{
\begin{array}{lr}
J_{1}^{(0)}=J_{2}^{(1)}J_{0}^{(1)}J_{1}^{(1)},\qquad \quad \ \mbox{on} \qquad (-\infty, k_{1}^{-}),\\
J_{1}^{(0)}=J_{4}^{(1)}J_{3}^{(1)},\qquad\qquad \quad \mbox{on} \qquad (k_{1}^{-}, \infty),\\
J_{2}^{(0)}=(J_{3-}^{(1)})^{-1}J_{\eta}^{(1)}J_{3+}^{(1)},\qquad \mbox{on} \qquad \eta_{+} \qquad \mbox{cut},\\
J_{3}^{(0)}=J_{4-}^{(1)}J_{\eta}^{(1)}(J_{4+}^{(1)})^{-1},\qquad \mbox{on} \qquad \eta_{-} \qquad \mbox{cut},\notag
\end{array}
\right.
\end{align}
where
\begin{gather}
J_{0}^{(1)}=\left(\begin{array}{cc}
   1+\gamma\gamma^{\ast}  &  0\\
   0 &  \frac{1}{1+\gamma\gamma^{\ast}}\\
\end{array}\right),\
J_{1}^{(1)}=\left(\begin{array}{cc}
   \Delta^{-\frac{1}{2}}  &  \frac{\Delta^{\frac{1}{2}}\gamma^{\ast}e^{2i\theta t}}{1+\gamma\gamma^{\ast}}\\
    0&  \Delta^{\frac{1}{2}}\\
\end{array}\right),\ J_{2}^{(1)}=\left(\begin{array}{cc}
   \Delta^{-\frac{1}{2}}  &  0\\
    \frac{\Delta^{\frac{1}{2}}\gamma e^{-2i\theta t}}{1+\gamma\gamma^{\ast}}&  \Delta^{\frac{1}{2}}\\
\end{array}\right),
\notag\\
J_{3}^{(1)}=\left(\begin{array}{cc}
   \Delta^{-\frac{1}{2}}  &   0\\
  \Delta^{-\frac{1}{2}}\gamma e^{-2i\theta t}  &  \Delta^{\frac{1}{2}}\\
\end{array}\right),\ J_{4}^{(1)}=\left(\begin{array}{cc}
   \Delta^{-\frac{1}{2}}  &   \Delta^{-\frac{1}{2}}\gamma^{\ast} e^{2i\theta t}\\
    0 &  \Delta^{\frac{1}{2}}\\
\end{array}\right),\ J_{\eta}^{(1)}=\left(\begin{array}{cc}
   0  &   \frac{q_{-}}{iq_{0}}\\
  \frac{q^{\ast}_{-}}{iq_{0}} &  0\\
\end{array}\right). \label{40}
\end{gather}
Defining a transformation
\begin{align}\label{41}
m^{(1)}=m^{(0)}B(k),
\end{align}
where
\begin{align}\label{42}
B(k)=\left\{
\begin{array}{lr}
(J_{1}^{(1)})^{-1}\ \mbox{on} \ k\in\Omega_{1},\\
J_{2}^{(1)}\ \mbox{on} \ k\in\Omega_{2},\\
(J_{3}^{(1)})^{-1}\ \mbox{on} \ k\in\Omega_{3}\cup\Omega_{5},\\
J_{4}^{(1)}\ \mbox{on} \ k\in\Omega_{4}\cup\Omega_{6},\\
I \ \mbox{on} \ k\in \mbox{others},
  \end{array}
\right.
\end{align}
we give out the following RHP about $m^{(1)}$.
\begin{align}\label{42.1}
\left\{
\begin{array}{lr}
m^{(1)}(x, t, k)\ \mbox{is analytic in} \ \mathbb{C}\setminus\Sigma^{(1)},\\
m^{(1)}_{+}(x, t, k)=m^{(1)}_{-}(x, t, k)J^{(1)}(x, t, k), \qquad k\in\Sigma^{(1)},\\
m^{(1)}(x, t, k)\rightarrow I,\qquad k\rightarrow \infty,
  \end{array}
\right.
\end{align}
of which the jump matrix $J^{(1)}$ is given in \eqref{40}.
This completes the deformation from the contour $\Sigma^{(0)}$  in Fig. 3 to
the  new contour $\Sigma^{(1)}$ of the new matrix-value function $m^{(1)}$ in Fig. 4.\\

\centerline{\begin{tikzpicture}
\draw[->][thick](-6.5,1)--(-6,1);
\draw [-,thick] (-6,1) to [out=-5,in=145] (-4,0);
\draw [->,thick] (-4,0) to [out=-35,in=180] (-3,-0.8);
\draw [-,thick] (-3,-0.8) to [out=0,in=-160] (0,0);
\draw [->,thick] (0,0) to [out=20,in=-175] (2,1);
\draw[-][thick](2,1)--(2.5,1);
\draw[->][thick](-6.5,-1)--(-6,-1);
\draw [-,thick] (-6,-1) to [out=5,in=-145] (-4,0);
\draw [->,thick] (-4,0) to [out=35,in=180] (-3,0.8);
\draw [-,thick] (-3,0.8) to [out=0,in=160] (0,0);
\draw [->,thick] (0,0) to [out=-20,in=175] (2,-1);
\draw[-][thick](2,-1)--(2.5,-1);
\draw[->][thick](-6.5,0)--(-5,0);
\draw[-][thick](-5,0)--(-4,0);
\draw[-][dashed,thick](-4,0)--(4,0);
\draw[->][very thick](0,0)--(0,1);
\draw[-][very thick](0,1)--(0,2);
\draw[-][very thick](0,0)--(0,-1);
\draw[<-][very thick](0,-1)--(0,-2);
\draw [->,thick] (0,0) to [out=130,in=-90] (-1,1);
\draw [-,thick]  (-1,1) to [out=90,in=-135] (0,2);
\draw [-,thick] (0,0) to [out=-130,in=90] (-1,-1);
\draw [<-,thick]  (-1,-1) to [out=-90,in=135] (0,-2);
\draw [->,thick] (0,0) to [out=50,in=-90] (1,1);
\draw [-,thick]  (1,1) to [out=90,in=-45] (0,2);
\draw [-,thick] (0,0) to [out=-50,in=90] (1,-1);
\draw [<-,thick]  (1,-1) to [out=-90,in=45] (0,-2);
\draw[fill] (-5,0.45) node[left]{$\Omega_{1}$};
\draw[fill] (-5,-0.45) node[left]{$\Omega_{2}$};
\draw[fill] (-3,0.45) node{$\Omega_{3}$};
\draw[fill] (-3,-0.45) node{$\Omega_{4}$};
\draw[fill] (2,0.45) node{$\Omega_{3}$};
\draw[fill] (2,-0.45) node{$\Omega_{4}$};
\draw[fill] (0,1.25) node[left]{$\Omega_{5}$};
\draw[fill] (0,-1.25) node[left]{$\Omega_{6}$};
\draw[fill] (0,1.25) node[right]{$\Omega_{5}$};
\draw[fill] (0,-1.25) node[right]{$\Omega_{6}$};
\draw[fill] (-4,-0.3) node{$k_{1}^{-}$};
\draw[fill] (-4,0) circle [radius=0.035];
\draw[fill] (0,-2) node[below]{$-iq_{0}-\frac{\beta}{2}$};
\draw[fill] (0,2) node[above]{$iq_{0}-\frac{\beta}{2}$};
\draw[fill] (-6,1) node[above]{$\textcolor[rgb]{1.00,0.00,0.00}{J_{1}^{(1)}}$};
\draw[fill] (-6,-1) node[below]{$\textcolor[rgb]{1.00,0.00,0.00}{J_{2}^{(1)}}$};
\draw[fill] (-6,0) node[above]{$\textcolor[rgb]{1.00,0.00,0.00}{J_{0}^{(1)}}$};
\draw[fill] (-3,0.8) node[above]{$\textcolor[rgb]{1.00,0.00,0.00}{J_{3}^{(1)}}$};
\draw[fill] (-3,-0.8) node[below]{$\textcolor[rgb]{1.00,0.00,0.00}{J_{4}^{(1)}}$};
\draw[fill] (2,1) node[above]{$\textcolor[rgb]{1.00,0.00,0.00}{J_{3}^{(1)}}$};
\draw[fill] (2,-1) node[below]{$\textcolor[rgb]{1.00,0.00,0.00}{J_{4}^{(1)}}$};
\draw[fill] (-0.5,1.75) node[left]{$\textcolor[rgb]{1.00,0.00,0.00}{J_{3}^{(1)}}$};
\draw[fill] (-0.5,-1.75) node[left]{$\textcolor[rgb]{1.00,0.00,0.00}{(J_{4}^{(1)})^{-1}}$};
\draw[fill] (0.5,1.75) node[right]{$\textcolor[rgb]{1.00,0.00,0.00}{(J_{3}^{(1)})^{-1}}$};
\draw[fill] (0.5,-1.75) node[right]{$\textcolor[rgb]{1.00,0.00,0.00}{J_{4}^{(1)}}$};
\draw[fill] (0,0.75) node[right]{$\textcolor[rgb]{1.00,0.00,0.00}{J_{\eta}^{(1)}}$};
\draw[fill] (0,-0.75) node[right]{$\textcolor[rgb]{1.00,0.00,0.00}{J_{\eta}^{(1)}}$};
\end{tikzpicture}}
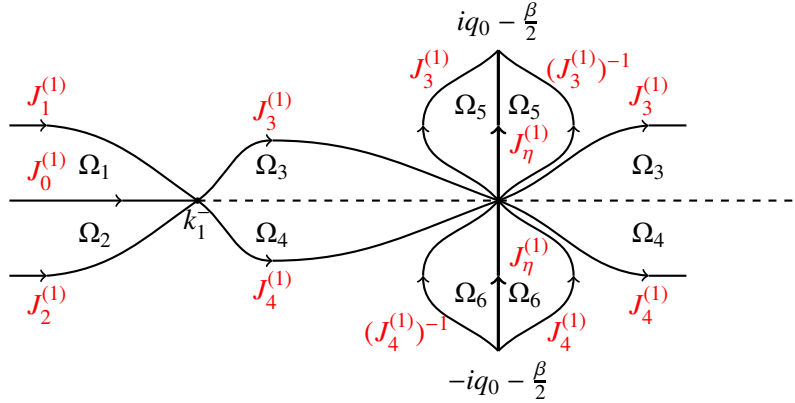
\centerline{\noindent {\small \textbf{Figure 3.} (Color online) The initial  contour  $\Sigma^{(0)}$.}}

\centerline{\begin{tikzpicture}
\draw[->][thick](-6.5,1)--(-6,1);
\draw [-,thick] (-6,1) to [out=-5,in=145] (-4,0);
\draw [->,thick] (-4,0) to [out=45,in=180] (-2,1.2);
\draw[-][thick](-2,1.2)--(0,1.2);
\draw [->,thick] (0,1.2) to [out=0,in=165] (1,0.3);
\draw [-,thick] (1,0.3) to [out=-15,in=180] (2,0.1);
\draw[->][thick](-6.5,-1)--(-6,-1);
\draw [-,thick] (-6,-1) to [out=5,in=-145] (-4,0);
\draw [->,thick] (-4,0) to [out=-45,in=-180] (-2,-1.2);
\draw[-][thick](-2,-1.2)--(0,-1.2);
\draw [->,thick] (0,-1.2) to [out=0,in=-165] (1,-0.3);
\draw [-,thick] (1,-0.3) to [out=15,in=-180] (2,-0.1);
\draw[->][thick](-6.5,0)--(-5,0);
\draw[-][thick](-5,0)--(-4,0);
\draw[->][very thick](-2,0)--(-2,0.4);
\draw[-][very thick](-2,0.4)--(-2,0.8);
\draw[-][very thick](-2,0)--(-2,-0.4);
\draw[-][very thick](-2,-0.4)--(-2,-0.8);
\draw[fill] (-4,-0.3) node{$k_{1}^{-}$};
\draw[fill] (-4,0) circle [radius=0.035];
\draw[fill] (-2,-0.8) node[right]{$-iq_{0}-\frac{\beta}{2}$};
\draw[fill] (-2,0.8) node[right]{$iq_{0}-\frac{\beta}{2}$};
\draw[fill] (-6,1) node[above]{$\textcolor[rgb]{1.00,0.00,0.00}{J_{1}^{(1)}}$};
\draw[fill] (-6,-1) node[below]{$\textcolor[rgb]{1.00,0.00,0.00}{J_{2}^{(1)}}$};
\draw[fill] (-6,0) node[above]{$\textcolor[rgb]{1.00,0.00,0.00}{J_{0}^{(1)}}$};
\draw[fill] (-2,0.4) node[left]{$\textcolor[rgb]{1.00,0.00,0.00}{J_{\eta}^{(1)}}$};
\draw[fill] (-2,1.2) node[above]{$\textcolor[rgb]{1.00,0.00,0.00}{J_{3}^{(1)}}$};
\draw[fill] (-2,-1.2) node[below]{$\textcolor[rgb]{1.00,0.00,0.00}{J_{4}^{(1)}}$};
\end{tikzpicture}}
\centerline{\noindent {\small \textbf{Figure 4.} (Color online) The  contour  $\Sigma^{(1)}$.}}
\subsection{Second deformation}
The second deformation is to remove the jump across the cut $(-\infty, k_{1}^{-})$. Thus, we introduce a scale RH problem, given by
\begin{align}\label{43}
\left\{
\begin{array}{lr}
\delta(k)\ \mbox{is analytic in} \ \mathbb{C}\setminus (-\infty, k_{1}^{-}),\\
\delta_{+}(k)=\delta_{-}(k)[1+\gamma(k)\gamma^{\ast}(k)],\qquad k\in (-\infty, k_{1}^{-}),\\
\delta(k)\rightarrow 1,\qquad k\rightarrow \infty.
  \end{array}
\right.
\end{align}
By Plemelj formula, the RH problem \eqref{43} has following solution
\begin{align}\label{45}
\delta(k)=\exp\big\{\frac{1}{2\pi i}\int_{-\infty}^{k_{1}^{-}}\frac{\ln[1+\gamma(y)\gamma^{\ast}(y)]}{y-k}dy\big\}.
\end{align}
Through transformation
\begin{align}\label{46}
m^{(2)}=m^{(1)}\delta^{-\sigma_{3}},
\end{align}
a new matrix-value function $m^{(2)}$ satisfies the following RH problem
\begin{align}\label{46.1}
\left\{
\begin{array}{lr}
m^{(2)}(x, t, k)\ \mbox{is analytic in} \ \mathbb{C}\setminus\Sigma^{(2)},\\
m^{(2)}_{+}(x, t, k)=m^{(2)}_{-}(x, t, k)J^{(2)}(x, t, k), \qquad k\in\Sigma^{(2)},\\
m^{(2)}(x, t, k)\rightarrow I,\qquad k\rightarrow \infty,
  \end{array}
\right.
\end{align}
where the contour $\Sigma^{(2)}$ is shown in Fig. 5, and $J^{(2)}=\delta_{-}^{\sigma_{3}}J^{(1)}\delta_{+}^{-\sigma_{3}}$, accurately given by
\begin{gather}
J_{1}^{(2)}=\left(\begin{array}{cc}
   \Delta^{-\frac{1}{2}}  &  \delta^{2}\frac{\Delta^{\frac{1}{2}}\gamma^{\ast}e^{2i\theta t}}{1+\gamma\gamma^{\ast}}\\
    0&  \Delta^{\frac{1}{2}}\\
\end{array}\right),\ J_{2}^{(2)}=\left(\begin{array}{cc}
   \Delta^{-\frac{1}{2}}  &  0\\
    \delta^{-2}\frac{\Delta^{\frac{1}{2}}\gamma e^{-2i\theta t}}{1+\gamma\gamma^{\ast}}&  \Delta^{\frac{1}{2}}\\
\end{array}\right),
\notag\\
J_{3}^{(2)}=\left(\begin{array}{cc}
   \Delta^{-\frac{1}{2}}  &   0\\
  \delta^{-2}\Delta^{-\frac{1}{2}}\gamma e^{-2i\theta t}  &  \Delta^{\frac{1}{2}}\\
\end{array}\right),\ J_{4}^{(2)}=\left(\begin{array}{cc}
   \Delta^{-\frac{1}{2}}  &   \delta^{2}\Delta^{-\frac{1}{2}}\gamma^{\ast} e^{2i\theta t}\\
    0 &  \Delta^{\frac{1}{2}}\\
\end{array}\right),\ J_{\eta}^{(2)}=\left(\begin{array}{cc}
   0  &   \delta^{2}\frac{q_{-}}{iq_{0}}\\
  \delta^{-2}\frac{q^{\ast}_{-}}{iq_{0}} &  0\\
\end{array}\right). \label{47}
\end{gather}

\centerline{\begin{tikzpicture}
\draw[->][thick](-6.5,1)--(-6,1);
\draw [-,thick] (-6,1) to [out=-5,in=145] (-4,0);
\draw [->,thick] (-4,0) to [out=45,in=180] (-2,1.2);
\draw[-][thick](-2,1.2)--(0,1.2);
\draw [->,thick] (0,1.2) to [out=0,in=165] (1,0.3);
\draw [-,thick] (1,0.3) to [out=-15,in=180] (2,0.1);
\draw[->][thick](-6.5,-1)--(-6,-1);
\draw [-,thick] (-6,-1) to [out=5,in=-145] (-4,0);
\draw [->,thick] (-4,0) to [out=-45,in=-180] (-2,-1.2);
\draw[-][thick](-2,-1.2)--(0,-1.2);
\draw [->,thick] (0,-1.2) to [out=0,in=-165] (1,-0.3);
\draw [-,thick] (1,-0.3) to [out=15,in=-180] (2,-0.1);
\draw[->][very thick](-2,0)--(-2,0.4);
\draw[-][very thick](-2,0.4)--(-2,0.8);
\draw[-][very thick](-2,0)--(-2,-0.4);
\draw[-][very thick](-2,-0.4)--(-2,-0.8);
\draw[fill] (-4,-0.3) node{$k_{1}^{-}$};
\draw[fill] (-4,0) circle [radius=0.035];
\draw[fill] (-2,-0.8) node[right]{$-iq_{0}-\frac{\beta}{2}$};
\draw[fill] (-2,0.8) node[right]{$iq_{0}-\frac{\beta}{2}$};
\draw[fill] (-6,1) node[above]{$\textcolor[rgb]{1.00,0.00,0.00}{J_{1}^{(2)}}$};
\draw[fill] (-6,-1) node[below]{$\textcolor[rgb]{1.00,0.00,0.00}{J_{2}^{(2)}}$};
\draw[fill] (-2,0.4) node[left]{$\textcolor[rgb]{1.00,0.00,0.00}{J_{\eta}^{(2)}}$};
\draw[fill] (-2,1.2) node[above]{$\textcolor[rgb]{1.00,0.00,0.00}{J_{3}^{(2)}}$};
\draw[fill] (-2,-1.2) node[below]{$\textcolor[rgb]{1.00,0.00,0.00}{J_{4}^{(2)}}$};
\draw[fill] (-5,0) node[left]{$\widehat{\Omega}_{3}$};
\draw[fill] (-4,0.5) node[above]{$\widehat{\Omega}_{1}$};
\draw[fill] (-4,-0.5) node[below]{$\widehat{\Omega}_{2}$};
\draw[fill] (-1,0) node{$\widehat{\Omega}_{4}$};
\end{tikzpicture}}
\centerline{\noindent {\small \textbf{Figure 5.} (Color online) The  contour  $\Sigma^{(2)}$.}}
\subsection{Third deformation}
The purpose of third deformation is to get rid of the term $\Delta(k)$.
Taking the following transformation
\begin{align}\label{48}
m^{(3)}=m^{(2)}\widehat{B}(k),
\end{align}
with
\begin{align}\label{49}
\widehat{B}(k)=\left\{
\begin{array}{lr}
\Delta^{\frac{\sigma_{3}}{2}}\ \mbox{on} \ k\in\widehat{\Omega}_{1},\\
\Delta^{-\frac{\sigma_{3}}{2}}\ \mbox{on} \ k\in\widehat{\Omega}_{2},\\
I \ \mbox{on} \ k\in \widehat{\Omega}_{3}\cup\widehat{\Omega}_{4},
  \end{array}
\right.
\end{align}
we obtain the following RH problem about $m^{(3)}$
\begin{align}\label{49.1}
\left\{
\begin{array}{lr}
m^{(3)}(x, t, k)\ \mbox{is analytic in} \ \mathbb{C}\setminus\Sigma^{(3)},\\
m^{(3)}_{+}(x, t, k)=m^{(3)}_{-}(x, t, k)J^{(3)}(x, t, k), \qquad k\in\Sigma^{(3)},\\
m^{(3)}(x, t, k)\rightarrow I,\qquad k\rightarrow \infty,
  \end{array}
\right.
\end{align}
where the contour $\Sigma^{(3)}=\Sigma^{(2)}$ is presented in Fig. 5, and $J^{(3)}$ is given by
\begin{gather}
J_{1}^{(3)}=\left(\begin{array}{cc}
   1  &  \delta^{2}\frac{\gamma^{\ast}e^{2i\theta t}}{1+\gamma\gamma^{\ast}}\\
    0&  1\\
\end{array}\right),\ J_{2}^{(3)}=\left(\begin{array}{cc}
   1  &  0\\
    \delta^{-2}\frac{\gamma e^{-2i\theta t}}{1+\gamma\gamma^{\ast}}&  1\\
\end{array}\right),
\notag\\
J_{3}^{(3)}=\left(\begin{array}{cc}
   1  &   0\\
  \delta^{-2}\gamma e^{-2i\theta t}  &  1\\
\end{array}\right),\ J_{4}^{(3)}=\left(\begin{array}{cc}
   1  &   \delta^{2}\gamma^{\ast} e^{2i\theta t}\\
    0 &  1\\
\end{array}\right),\ J_{\eta}^{(3)}=J_{\eta}^{(2)}. \label{50}
\end{gather}
\subsection{The $g$-function and model problem}
In order to make the above RH problem \eqref{49.1} become the solvable RH problem, we define a transformation
\begin{align}\label{51}
m^{(4)}=m^{(3)}e^{ig(k)\sigma_{3}},
\end{align}
of which $g(k)$ is analytic in $\mathbb{C}\setminus \eta$, and the following discontinuity condition need to be satisfied
\begin{align}\label{52}
\delta^{-2}(k)e^{i(g_{+}(k)+g_{-}(k))}=1,\qquad k\in \eta.
\end{align}
After a simple calculation, then the jump matrix $J_{\eta}^{(4)}$ across $\eta$ is changed into
\begin{align}\label{53}
J_{\eta}^{(4)}=\left(\begin{array}{cc}
   0  &   \frac{q_{-}}{iq_{0}}\\
  \frac{q^{\ast}_{-}}{iq_{0}} &  0\\
\end{array}\right),\qquad k\in \eta,
\end{align}
which is exactly a constant, and the
other jump matrices in Eq.\eqref{50}  turn into
\begin{gather}
J_{1}^{(4)}=\left(\begin{array}{cc}
   1  &  \delta^{2}\frac{\gamma^{\ast}e^{2i(\theta t-g)}}{1+\gamma\gamma^{\ast}}\\
    0&  1\\
\end{array}\right),\ J_{2}^{(4)}=\left(\begin{array}{cc}
   1  &  0\\
    \delta^{-2}\frac{\gamma e^{-2i(\theta t-g)}}{1+\gamma\gamma^{\ast}}&  1\\
\end{array}\right),
\notag\\
J_{3}^{(4)}=\left(\begin{array}{cc}
   1  &   0\\
  \delta^{-2}\gamma e^{-2i(\theta t-g)}  &  1\\
\end{array}\right),\ J_{4}^{(4)}=\left(\begin{array}{cc}
   1  &   \delta^{2}\gamma^{\ast} e^{2i(\theta t-g)}\\
    0 &  1\\
\end{array}\right). \label{54}
\end{gather}
Considering $\lambda_{-}=-\lambda_{+}$ on the $\eta$ cut and using the Plemelj's formula, one gets
\begin{align}\label{54.1}
g(k)=\frac{\lambda(k)}{2\pi^{2}i}\int_{z\in \eta}\frac{1}{\lambda(z)(z-k)}\int_{-\infty}^{k_{1}^{-}}\frac{\ln[1+\gamma(y)
\gamma^{\ast}(y)]}{y-z}dydz.
\end{align}
Making $k\rightarrow\infty$ in \eqref{54.1}, we obtain
\begin{align}\label{55}
g(\infty)=-\frac{1}{2\pi^{2}i}\int_{z\in \eta}\frac{1}{\lambda(z)}\int_{-\infty}^{k_{1}^{-}}\frac{\ln[1+\gamma(y)\gamma^{\ast}(y)]}{y-z}dydz,
\end{align}
which is a function of the $q_{0}, \beta$ and $\xi$, whereas is independent of $k$. As $t\rightarrow\infty$, the jump matrices $J_{i}^{(4)}, (i = 1, 2, 3, 4)$ will be degenerative exponentially to the identity far from the
point $k_{1}^{-}$. Finally, $m^{(4)}$ can be expressed into the form
\begin{align}\label{56}
m^{(4)}=m^{err}m^{mod},
\end{align}
where $m^{err}$ problem yields
\begin{align}\label{57}
m^{err}=I+\mathcal{O}(t^{-\frac{1}{2}}),
\end{align}
but the model problem will  dominate the long-time asymptotics of solution $q(x,t)$. Let $m^{mod}$ solve the RH problem:
\begin{align}\label{58}
\left\{
\begin{array}{lr}
m^{mod}(x, t, k)\ \mbox{is analytic in} \ \mathbb{C}\setminus \eta,\\
m^{mod}_{+}(x, t, k)=m^{mod}_{-}(x, t, k)J^{mod}(x, t, k), \qquad k\in\eta,\\
m^{mod}(x, t, k)\rightarrow e^{ig(\infty)\sigma_{3}},\qquad k\rightarrow \infty,
  \end{array}
\right.
\end{align}
of which the jump matrix $J^{mod}(x, t, k)=J_{\eta}^{(4)}(x, t, k)$ given in \eqref{53}, and $m^{mod}$ can be expressed exactly as
\begin{align}\label{59}
m^{mod}=e^{ig(\infty)\sigma_{3}}\left(\begin{array}{cc}
   \frac{1}{2}(\Upsilon+\Upsilon^{-1})  &   -\frac{q_{0}}{2q^{\ast}_{-}}(\Upsilon-\Upsilon^{-1})\\
  \frac{q_{0}}{2q_{-}}(\Upsilon-\Upsilon^{-1})^{\ast}  &  \frac{1}{2}(\Upsilon+\Upsilon^{-1})^{\ast}\\
\end{array}\right),
\end{align}
where
\begin{align}\label{60}
\Upsilon(k)=\left(\frac{k-iq_{0}+\frac{\beta}{2}}{k+iq_{0}+\frac{\beta}{2}}\right)^{\frac{1}{4}}.
\end{align}
Then, going back to the formula for potential $q(x, t)$ in \eqref{34}, we have
\begin{align}\label{61}
q(x,t)&=-2i\left(m_{1}^{(0)}(x,t)\right)_{12}e^{i[-\beta x+(\alpha-2q_{0}^{2})t]}\notag\\
&=-2i\left(m_{1}^{(3)}(x,t)\right)_{12}e^{i[-\beta x+(\alpha-2q_{0}^{2})t]}\notag\\
&=-2i\left(m_{1}^{mod}(x,t)e^{-ig(\infty)\sigma_{3}}\right)_{12}e^{i[-\beta x+(\alpha-2q_{0}^{2})t]}+\mathcal{O}(t^{-\frac{1}{2}}),
\end{align}
where $m_{1}^{mod}$ is given by the explicit solution $m^{mod}$ in Eq. \eqref{59}
\begin{align}\label{62}
m_{1}^{mod}&=\lim\limits_{k\rightarrow \infty}k(m^{mod}-e^{ig(\infty)\sigma_{3}})\notag\\
&=e^{ig(\infty)\sigma_{3}}\lim\limits_{k\rightarrow \infty}k\left(\begin{array}{cc}
   \frac{1}{2}(\Upsilon+\Upsilon^{-1})-1  &   -\frac{q_{0}}{2q^{\ast}_{-}}(\Upsilon-\Upsilon^{-1})\\
  \frac{q_{0}}{2q_{-}}(\Upsilon-\Upsilon^{-1})^{\ast}  &  \frac{1}{2}(\Upsilon+\Upsilon^{-1})^{\ast}-1\\
\end{array}\right).
\end{align}
Note that
\begin{align}\label{63}
\Upsilon-\Upsilon^{-1}=\frac{-iq_{0}}{k}+O(\frac{1}{k}),\qquad \mbox{as} \quad k\rightarrow\infty,
\end{align}
then we have
\begin{align}\label{64}
\left(m_{1}^{mod}(x,t)e^{-ig(\infty)\sigma_{3}}\right)_{12}=-e^{2ig(\infty)}\lim\limits_{k\rightarrow \infty}k\frac{q_{0}}{2q^{\ast}_{-}}(\Upsilon-\Upsilon^{-1})=\frac{iq_{-}}{2}e^{2ig(\infty)}.
\end{align}
Thus, in the plane wave region I, we can declare that the long-time asymptotic behavior of the solution for the Eq.\eqref{3.1}  is given by
\begin{align}\label{65}
q(x,t)=q_{-}e^{i[2g(\infty)-\beta x+(\alpha-2q_{0}^{2})t]}+\mathcal{O}(t^{-\frac{1}{2}}),
\end{align}
where $g(\infty)$ is shown in Eq.\eqref{55}, which is only dependent of $q_{0}, \beta$ and $\xi$ with $\xi>\xi_{2}$.

\section{Modulated elliptic wave region}
For the modulated elliptic wave region $0<\xi<\xi_{2}$, the curves $\mbox{Im}\theta(k)=0$ will not intersect the real axis.  In order to study the long-time asymptotics of $q(x,t)$ in this region, we introduce a $g$-function mechanism and perform the same first, second and third deformations as that done in the plane wave region I, but
the change of factorization  happens at the point $k_{1}^{-}$ shall be replaced by the point $k_{1}^{0}$. Then we obtain the RH problem \eqref{49.1}. Moreover, the further deformations
for the contour $\Sigma^{(3)}$ still need to be considered until  the solvable RH problem is derived.
\subsection{Eliminating of the exponential growth}
When $0<\xi<\xi_{2}$ and as $t$ tends to infinity, the jump matrices $J_{3}^{(3)}$ and $J_{4}^{(3)}$ in Eq. \eqref{50} will  grow exponentially in the segment $[k_{1}^{0},\chi]$ and the segment $[k_{1}^{0},\chi^{\ast}]$, respectively. To solve this barrier, we decompose matrices $J_{3}^{(3)}$ and $J_{4}^{(3)}$ into  following form (see Fig. 6)
\begin{align}\label{67}
J_{3}^{(3)}=J_{5}^{(3)}J_{7}^{(3)}J_{5}^{(3)},\qquad J_{4}^{(3)}=J_{6}^{(3)}J_{8}^{(3)}J_{6}^{(3)},
\end{align}
where
\begin{gather}\label{68}
J_{5}^{(3)}=\left(\begin{array}{cc}
   1  &  \delta^{2}\gamma^{-1}e^{2i\theta t}\\
   0 &  1\\
\end{array}\right),\quad J_{6}^{(3)}=\left(\begin{array}{cc}
   1  &  0\\
   \delta^{-2}(\gamma^{\ast})^{-1}e^{-2i\theta t} &  1\\
\end{array}\right),\notag\\
J_{7}^{(3)}=\left(\begin{array}{cc}
   0  &  -\delta^{2}\gamma^{-1}e^{2i\theta t}\\
  \delta^{-2}\gamma e^{-2i\theta t} &  0\\
\end{array}\right),\quad J_{8}^{(3)}=\left(\begin{array}{cc}
   0  &  \delta^{2}\gamma^{\ast}e^{2i\theta t}\\
   -\delta^{-2}(\gamma^{\ast})^{-1}e^{-2i\theta t} &  0\\
\end{array}\right).
\end{gather}

\centerline{\begin{tikzpicture}[scale=1.2]
\draw[->][thick](-6.5,1)--(-6,1);
\draw [-,thick] (-6,1) to [out=-5,in=145] (-4,0);
\draw[->][thick] (-4,0)--(-3.5,0.5);
\draw [->,thick] (-3.5,0.5) to [out=45,in=180] (-2,1.2);
\draw [->,thick] (-4,0) to [out=90,in=-135] (-3.75,0.8);
\draw [-,thick] (-3.75,0.8) to [out=45,in=180] (-3,0.9);
\draw [->,thick] (-4,0) to [out=0,in=-135] (-3.25,0.2);
\draw [-,thick] (-3.25,0.2) to [out=45,in=-90] (-3,0.9);
\draw[fill] (-3,0.9) node[above]{$\chi$} circle [radius=0.035];
\draw[fill] (-3,-0.9) node[below]{$\chi^{\ast}$} circle [radius=0.035];
\draw [->,thick] (-4,0) to [out=-90,in=135] (-3.75,-0.8);
\draw [-,thick] (-3.75,-0.8) to [out=-45,in=180] (-3,-0.9);
\draw [->,thick] (-4,0) to [out=0,in=135] (-3.25,-0.2);
\draw [-,thick] (-3.25,-0.2) to [out=-45,in=90] (-3,-0.9);
\draw[-][thick](-2,1.2)--(0,1.2);
\draw [->,thick] (0,1.2) to [out=0,in=165] (1,0.3);
\draw [-,thick] (1,0.3) to [out=-15,in=180] (2,0.1);
\draw[->][thick](-6.5,-1)--(-6,-1);
\draw [-,thick] (-6,-1) to [out=5,in=-145] (-4,0);
\draw[->][thick] (-4,0)--(-3.5,-0.5);
\draw [->,thick] (-3.5,-0.5) to [out=-45,in=-180] (-2,-1.2);
\draw[-][thick](-2,-1.2)--(0,-1.2);
\draw [->,thick] (0,-1.2) to [out=0,in=-165] (1,-0.3);
\draw [-,thick] (1,-0.3) to [out=15,in=-180] (2,-0.1);
\draw[->][very thick](-2,0)--(-2,0.4);
\draw[-][very thick](-2,0.4)--(-2,0.8);
\draw[-][very thick](-2,0)--(-2,-0.4);
\draw[-][very thick](-2,-0.4)--(-2,-0.8);
\draw[fill] (-4.15,0) node[left]{$k_{1}^{0}$};
\draw[fill] (-4,0) circle [radius=0.035];
\draw[fill] (-2,-0.8) node[right]{$-iq_{0}-\frac{\beta}{2}$};
\draw[fill] (-2,0.8) node[right]{$iq_{0}-\frac{\beta}{2}$};
\draw[fill] (-6,1) node[above]{$\textcolor[rgb]{1.00,0.00,0.00}{J_{1}^{(3)}}$};
\draw[fill] (-6,-1) node[below]{$\textcolor[rgb]{1.00,0.00,0.00}{J_{2}^{(3)}}$};
\draw[fill] (-2,0) node[right]{$\textcolor[rgb]{1.00,0.00,0.00}{J_{\eta}^{(3)}}$};
\draw[fill] (-2,1.2) node[above]{$\textcolor[rgb]{1.00,0.00,0.00}{J_{3}^{(3)}}$};
\draw[fill] (-2,-1.2) node[below]{$\textcolor[rgb]{1.00,0.00,0.00}{J_{4}^{(3)}}$};
\draw[fill] (-3.75,0.8)node[above]{$\textcolor[rgb]{1.00,0.00,0.00}{J_{5}^{(3)}}$};
\draw[fill] (-3.2,0.3)node[right]{$\textcolor[rgb]{1.00,0.00,0.00}{J_{5}^{(3)}}$};
\draw[fill] (-3.3,0.5)node{$\textcolor[rgb]{1.00,0.00,0.00}{J_{7}^{(3)}}$};
\draw[fill] (-3.75,-0.8)node[below]{$\textcolor[rgb]{1.00,0.00,0.00}{J_{6}^{(3)}}$};
\draw[fill] (-3.2,-0.3)node[right]{$\textcolor[rgb]{1.00,0.00,0.00}{J_{6}^{(3)}}$};
\draw[fill] (-3.3,-0.5)node{$\textcolor[rgb]{1.00,0.00,0.00}{J_{8}^{(3)}}$};
\end{tikzpicture}}
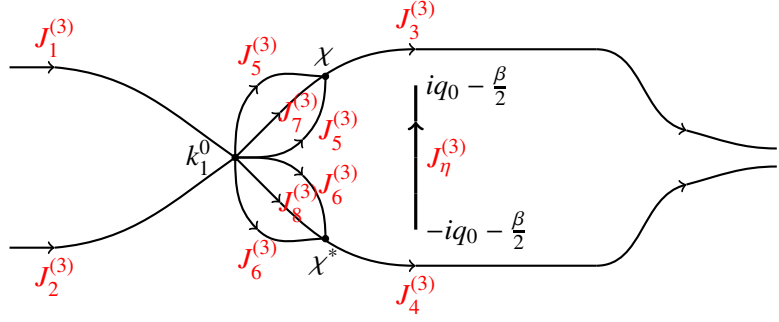
\centerline{\noindent {\small \textbf{Figure 6.} (Color online) The contour  $\widehat{\Sigma}^{(3)}$ for the modulated genus 1 elliptic wave sector.}}

Next, we employ a time-dependent $g$-function to make  the transformation
\begin{align}\label{69}
m^{(4)}=m^{(3)}e^{-iG(k)t\sigma_{3}},
\end{align}
where function $G(k)$ is analytic off the cuts $\eta\cup\varpi$, of which $\varpi=\varpi_{+}\cup\varpi_{-}$ with $\varpi_{+}=[k_{1}^{0},\chi]$ and
$\varpi_{-}=[k_{1}^{0},\chi^{\ast}]$. Then we can obtain the following new jump matrices $J^{(4)}$ whose jump  contour is $\widehat{\Sigma}^{(4)}$ (see Fig. 7)
\begin{gather}
J_{1}^{(4)}=\left(\begin{array}{cc}
   1  &  \delta^{2}\frac{\gamma^{\ast}e^{2i(\theta +G)t}}{1+\gamma\gamma^{\ast}}\\
    0&  1\\
\end{array}\right),\ J_{2}^{(4)}=\left(\begin{array}{cc}
   1  &  0\\
    \delta^{-2}\frac{\gamma e^{-2i(\theta +G)t}}{1+\gamma\gamma^{\ast}}&  1\\
\end{array}\right),
\notag\\
J_{3}^{(4)}=\left(\begin{array}{cc}
   1  &   0\\
  \delta^{-2}\gamma e^{-2i(\theta +G)t}  &  1\\
\end{array}\right),\ J_{4}^{(4)}=\left(\begin{array}{cc}
   1  &   \delta^{2}\gamma^{\ast} e^{2i(\theta +G)t}\\
    0 &  1\\
\end{array}\right),\notag\\
\ J_{\eta}^{(4)}=\left(\begin{array}{cc}
   0  &   \delta^{2}\frac{q_{-}}{iq_{0}}e^{i(G_{+}+G_{-})t}\\
  \delta^{-2}\frac{q^{\ast}_{-}}{iq_{0}}e^{-i(G_{+}+G_{-})t} &  0\\
\end{array}\right),\notag\\
J_{5}^{(4)}=\left(\begin{array}{cc}
   1  &  \delta^{2}\gamma^{-1}e^{2i(\theta +G)t}\\
   0 &  1\\
\end{array}\right),\quad J_{6}^{(4)}=\left(\begin{array}{cc}
   1  &  0\\
   \delta^{-2}(\gamma^{\ast})^{-1}e^{-2i(\theta +G)t} &  1\\
\end{array}\right),\notag\\
J_{7}^{(4)}=\left(\begin{array}{cc}
   0  &  -\delta^{2}\gamma^{-1}e^{i(2\theta+G_{+}+G_{-})t}\\
  \delta^{-2}\gamma e^{-i(2\theta+G_{+}+G_{-})t} &  0\\
\end{array}\right),\notag\\ J_{8}^{(4)}=\left(\begin{array}{cc}
   0  &  \delta^{2}\gamma^{\ast}e^{i(2\theta+G_{+}+G_{-})t}\\
   -\delta^{-2}(\gamma^{\ast})^{-1}e^{-i(2\theta+G_{+}+G_{-})t} &  0\\
\end{array}\right). \label{70}
\end{gather}

\centerline{\begin{tikzpicture}[scale=1.2]
\draw[->][thick](-6.5,1)--(-6,1);
\draw [-,thick] (-6,1) to [out=-5,in=145] (-4,0);
\draw[->][thick] (-4,0)--(-3.5,0.5);
\draw [->,thick] (-3.5,0.5) to [out=45,in=180] (-2,1.2);
\draw [->,thick] (-4,0) to [out=90,in=-135] (-3.75,0.8);
\draw [-,thick] (-3.75,0.8) to [out=45,in=180] (-3,0.9);
\draw [->,thick] (-4,0) to [out=0,in=-135] (-3.25,0.2);
\draw [-,thick] (-3.25,0.2) to [out=45,in=-90] (-3,0.9);
\draw[fill] (-3,0.9) node[above]{$\chi$} circle [radius=0.035];
\draw[fill] (-3,-0.9) node[below]{$\chi^{\ast}$} circle [radius=0.035];
\draw [->,thick] (-4,0) to [out=-90,in=135] (-3.75,-0.8);
\draw [-,thick] (-3.75,-0.8) to [out=-45,in=180] (-3,-0.9);
\draw [->,thick] (-4,0) to [out=0,in=135] (-3.25,-0.2);
\draw [-,thick] (-3.25,-0.2) to [out=-45,in=90] (-3,-0.9);
\draw[-][thick](-2,1.2)--(0,1.2);
\draw [->,thick] (0,1.2) to [out=0,in=165] (1,0.3);
\draw [-,thick] (1,0.3) to [out=-15,in=180] (2,0.1);
\draw[->][thick](-6.5,-1)--(-6,-1);
\draw [-,thick] (-6,-1) to [out=5,in=-145] (-4,0);
\draw[->][thick] (-4,0)--(-3.5,-0.5);
\draw [->,thick] (-3.5,-0.5) to [out=-45,in=-180] (-2,-1.2);
\draw[-][thick](-2,-1.2)--(0,-1.2);
\draw [->,thick] (0,-1.2) to [out=0,in=-165] (1,-0.3);
\draw [-,thick] (1,-0.3) to [out=15,in=-180] (2,-0.1);
\draw[->][very thick](-2,0)--(-2,0.4);
\draw[-][very thick](-2,0.4)--(-2,0.8);
\draw[-][very thick](-2,0)--(-2,-0.4);
\draw[-][very thick](-2,-0.4)--(-2,-0.8);
\draw[fill] (-4.15,0) node[left]{$k_{1}^{0}$};
\draw[fill] (-4,0) circle [radius=0.035];
\draw[fill] (-2,-0.8) node[right]{$-iq_{0}-\frac{\beta}{2}$};
\draw[fill] (-2,0.8) node[right]{$iq_{0}-\frac{\beta}{2}$};
\draw[fill] (-6,1) node[above]{$\textcolor[rgb]{1.00,0.00,0.00}{J_{1}^{(4)}}$};
\draw[fill] (-6,-1) node[below]{$\textcolor[rgb]{1.00,0.00,0.00}{J_{2}^{(4)}}$};
\draw[fill] (-2,0) node[right]{$\textcolor[rgb]{1.00,0.00,0.00}{J_{\eta}^{(4)}}$};
\draw[fill] (-2,1.2) node[above]{$\textcolor[rgb]{1.00,0.00,0.00}{J_{3}^{(4)}}$};
\draw[fill] (-2,-1.2) node[below]{$\textcolor[rgb]{1.00,0.00,0.00}{J_{4}^{(4)}}$};
\draw[fill] (-3.75,0.8)node[above]{$\textcolor[rgb]{1.00,0.00,0.00}{J_{5}^{(4)}}$};
\draw[fill] (-3.2,0.3)node[right]{$\textcolor[rgb]{1.00,0.00,0.00}{J_{5}^{(4)}}$};
\draw[fill] (-3.3,0.5)node{$\textcolor[rgb]{1.00,0.00,0.00}{J_{7}^{(4)}}$};
\draw[fill] (-3.75,-0.8)node[below]{$\textcolor[rgb]{1.00,0.00,0.00}{J_{6}^{(4)}}$};
\draw[fill] (-3.2,-0.3)node[right]{$\textcolor[rgb]{1.00,0.00,0.00}{J_{6}^{(4)}}$};
\draw[fill] (-3.3,-0.5)node{$\textcolor[rgb]{1.00,0.00,0.00}{J_{8}^{(4)}}$};
\draw[-][dashed,thick](-6.5,0)--(2,0);
\draw[-] [dashed,thick] (-2,0.8) to [out=0,in=-45] (-3,0.9);
\draw[-] [dashed,thick] (-3,0.9) to [out=135,in=-90] (-3.8,2);
\draw[-] [dashed,thick] (-2,-0.8) to [out=0,in=45] (-3,-0.9);
\draw[-] [dashed,thick] (-3,-0.9) to [out=-135,in=90] (-3.8,-2);
\draw[fill] (-5,0.3)node{$\textbf{--}\ \textbf{--}$};
\draw[fill] (-5,-0.3)node{$\textbf{++}$};
\draw[fill] (-4.5,1)node{$\textbf{--}\ \textbf{--}$};
\draw[fill] (-4.5,-1)node{$\textbf{++}$};
\draw[fill] (-3,1.5)node{$\textbf{++}$};
\draw[fill] (-3,-1.5)node{$\textbf{--}\ \textbf{--}$};
\draw[fill] (-2.5,0.2)node{$\textbf{--}\ \textbf{--}$};
\draw[fill] (-2.5,-0.4)node{$\textbf{++}$};
\draw[fill] (0,0.4)node{$\textbf{++}$};
\draw[fill] (0,-0.4)node{$\textbf{--}\ \textbf{--}$};
\end{tikzpicture}}
\centerline{\noindent {\small \textbf{Figure 7.} (Color online) The contour  $\widehat{\Sigma}^{(4)}$ for the modulated genus 1 elliptic wave sector.}}

As a matter of convenience, we define a function $\omega$ as follows
\begin{align}\label{71}
\omega(k)=\theta(k)+G(k).
\end{align}
In order to give out the parameters $k_{1}^{0}$ and  $\chi$,  the properties of $\omega(k)$ need to be discussed. Firstly, we introduce a function $z$, given by
\begin{align}\label{72}
z(k)=\sqrt{[(k+\frac{\beta}{2})^{2}+q_{0}^{2}](k-\chi)(k-\chi^{\ast})},
\end{align}
which admits branch cuts $\eta\cup\varpi$. As well as, we set $z(k)=-z_{+}(k)=z_{-}(k)$.

Let $\omega(k)$ admits the following Abelian integral
\begin{align}\label{73}
\omega(k)=\frac{1}{2}\left(\int_{\frac{iq_{0}}{2}-\frac{\beta}{2}}^{k}
+\int_{\frac{-iq_{0}}{2}-\frac{\beta}{2}}^{k}\right)d\omega(y),
\end{align}
where the Abelian differential $d\omega$ is given by
\begin{align}\label{74}
d\omega(k)=4\frac{(k-k_{1}^{0})(k-\chi)(k-\chi^{\ast})}{z(k)}dk.
\end{align}
Taking $\chi=\chi_{1}+\chi_{2}i$,  we can easily get
\begin{align}\label{75}
(k-k_{1}^{0})(k-\chi)(k-\chi^{\ast})=k^{3}+h_{2}k^{2}+h_{1}k+h_{0},
\end{align}
where
\begin{align}\label{76}
h_{2}=-(k_{1}^{0}+2\chi_{1}),\qquad h_{1}=\chi_{1}^{2}+2k_{1}^{0}\chi_{1}+\chi_{2}^{2},\qquad h_{0}=-k_{1}^{0}(\chi_{1}^{2}+\chi_{2}^{2}).
\end{align}
Naturally, we have
\begin{align}\label{77}
\omega(k)=2\left(\int_{\frac{iq_{0}}{2}-\frac{\beta}{2}}^{k}
+\int_{\frac{-iq_{0}}{2}-\frac{\beta}{2}}^{k}\right)\frac{y^{3}+h_{2}y^{2}+h_{1}y+h_{0}}{z(y)}dy.
\end{align}
We suppose that the sign signatures of Im$\omega(k)$ is same as that of Im$\theta(k)$ for large $k$,
\begin{align}\label{78}
\mbox{Im} \omega=\mbox{Im} \theta+\mathcal{O}\frac{1}{k},\qquad k\rightarrow\infty.
\end{align}
 As $k\rightarrow\infty$, the expression of $z(k)$ become
\begin{align}\label{79}
z(k)=k(k+\frac{\beta}{2})\left[1-\frac{\chi+\chi^{\ast}}{2k}
+\frac{4q_{0}^{2}-(\chi-\chi^{\ast})^{2}}{8k^{2}}+\mathcal{O}(\frac{1}{k^{3}})\right],\quad k\rightarrow\infty,
\end{align}
which implies that
\begin{align}\label{80}
\frac{d\omega}{dk}=4\left(k+h_{2}+\chi_{1}-\frac{\beta}{2}+\frac{\Xi}{k}+\mathcal{O}(\frac{1}{k^{2}})\right), \quad k\rightarrow\infty,
\end{align}
where $\Xi=\frac{\beta^{2}}{4}-\frac{q_{0}^{2}}{2}
-\frac{\chi_{2}^{2}}{2}+h_{1}+(h_{2}+\chi_{1})(\chi_{1}-\frac{\beta}{2})$. After an integration, we
derive
\begin{align}\label{81}
\omega(k)=2k^{2}+4(h_{2}+\chi_{1}-\frac{\beta}{2})k+4\Xi\ln(k)+\omega_{0}+\mathcal{O}(\frac{1}{k}), \quad k\rightarrow\infty,
\end{align}
where $\omega_{0}$ is a constant to be known later. On the other hand, since $\theta=\lambda(\xi+2k)$, we obtain
\begin{align}\label{82}
\theta(k)=2k^{2}+(\xi+\beta)k+\frac{1}{2}\beta\xi+q_{0}^{2}+\mathcal{O}(\frac{1}{k}), \quad k\rightarrow\infty.
\end{align}
Since \eqref{78} is allowed, and from Eqs. \eqref{81}, \eqref{82}, we can derive $\Xi=0$
and
\begin{align}\label{83}
h_{2}=\frac{\xi+3\beta}{4}-\chi_{1},\qquad h_{1}=-\frac{\beta^{2}}{4}+\frac{q_{0}^{2}}{2}
+\frac{\chi_{2}^{2}}{2}-(\frac{\xi+3\beta}{4})(\chi_{1}-\frac{\beta}{2}),
\end{align}
which indicates that
\begin{align}\label{84}
\chi_{1}=-k_{1}^{0}-\frac{\xi+3\beta}{4},\qquad \chi_{2}=\sqrt{\frac{\xi k_{1}^{0}}{2}+\frac{\xi\beta}{4}+\frac{3\beta k_{1}^{0}}{2}+\frac{\beta^{2}}{4}+2(k_{1}^{0})^{2}+q_{0}^{2}},
\end{align}
where parameter $k_{1}^{0}$ is still known later. Observing that
\begin{align}\label{85}
2\left(\int_{\frac{iq_{0}}{2}-\frac{\beta}{2}}^{k}
+\int_{\frac{-iq_{0}}{2}-\frac{\beta}{2}}^{k}\right)(y+\frac{\xi+\beta}{4})dy
=2k^{2}+(\xi+\beta)k-\frac{\xi\beta+q_{0}^{2}}{4},
\end{align}
then we can write the expression of $\omega(k)$ in Eq.\eqref{77} into
\begin{align}\label{86}
\omega(k)&=2\left(\int_{\frac{iq_{0}}{2}-\frac{\beta}{2}}^{k}
+\int_{\frac{-iq_{0}}{2}-\frac{\beta}{2}}^{k}\right)\left[\frac{y^{3}+h_{2}y^{2}+h_{1}y+h_{0}}{z(y)}
-(y+\frac{\xi+\beta}{4})\right]dy\notag\\
&+2k^{2}+(\xi+\beta)k-\frac{\xi\beta+q_{0}^{2}}{4}.
\end{align}
Let $k\rightarrow\infty$ in Eqs.\eqref{81} and \eqref{86}, one has
\begin{align}\label{87}
\omega_{0}&=2\left(\int_{\frac{iq_{0}}{2}-\frac{\beta}{2}}^{\infty}
+\int_{\frac{-iq_{0}}{2}-\frac{\beta}{2}}^{\infty}\right)\left[\frac{y^{3}+h_{2}y^{2}+h_{1}y+h_{0}}{z(y)}
-(y+\frac{\xi+\beta}{4})\right]dy-\frac{\xi\beta+q_{0}^{2}}{4}.
\end{align}
Due to the following  the large-$k$ asymptotics
\begin{align}\label{88}
\frac{y^{3}+h_{2}y^{2}+h_{1}y+h_{0}}{z(y)}
-(y+\frac{\xi+\beta}{4})=\mathcal{O}(\frac{1}{k^{2}}),\quad k\rightarrow\infty,
\end{align}
Eq.\eqref{87} can be well-defined.

Next, we will  give out the parameter $k_{1}^{0}$ on the real line. In what follows, to make the jump matrices $J_{3}^{(4)}, J_{5}^{(4)}$ and $J_{6}^{(4)}$ be bounded, we
discuss the sign signature of Im$\omega(k)$ near point $\chi=\chi_{1}+\chi_{2}i$. We first present the asymptotic expansions of $\lambda(k)$ and $\sqrt{(k-\chi)(k-\chi^{\ast})}$ near point $\chi$, given by
\begin{align}\label{89}
\sqrt{(k+\frac{\beta}{2})^{2}+q_{0}^{2}}=\sqrt{(\chi+\frac{\beta}{2})^{2}+q_{0}^{2}}
\left[1+\frac{(\chi+\frac{\beta}{2})(k-\chi)}{(\chi+\frac{\beta}{2})^{2}+q_{0}^{2}}
+\mathcal{O}((k-\chi)^{2})\right],\qquad k\rightarrow\chi,
\end{align}
\begin{align}\label{90}
\sqrt{(k-\chi)(k-\chi^{\ast})}=\sqrt{k-\chi^{\ast}}\left[\sqrt{k-\chi}
+\frac{(k-\chi)^{\frac{3}{2}}}{2(\chi-\chi^{\ast})}+\mathcal{O}((k-\chi)^{\frac{5}{2}})\right],\qquad k\rightarrow\chi.
\end{align}
In the same way, we show the asymptotic behaviors of Eq.\eqref{74} near point $\chi$
\begin{gather}
\frac{d\omega}{dk}=\frac{8(\chi-k_{1}^{0})\sqrt{\chi-\chi^{\ast}}}{\sqrt{(\beta+2\chi)^{2}+4q_{0}^{2}}}
\left[(k-\chi)^{\frac{1}{2}}\right.\notag\\
\left.+[\frac{1}{\chi-k_{1}^{0}}+\frac{1}{2(\chi-\chi^{\ast})}
-\frac{2(\beta+2\chi)}{(\beta+2\chi)^{2}+4q_{0}^{2}}](k-\chi)^{\frac{3}{2}}+\mathcal{O}((k-\chi)^{\frac{5}{2}})\right],\qquad k\rightarrow\chi.\label{91}
\end{gather}
Further, one gets
\begin{gather}
\omega(k)=\omega(\chi)+\frac{8(\chi-k_{1}^{0})\sqrt{\chi-\chi^{\ast}}}{\sqrt{(\beta+2\chi)^{2}+4q_{0}^{2}}}
\left[\frac{2}{3}(k-\chi)^{\frac{3}{2}}\right.\notag\\
\left.+\frac{2}{5}[\frac{1}{\chi-k_{1}^{0}}+\frac{1}{2(\chi-\chi^{\ast})}
-\frac{2(\beta+2\chi)}{(\beta+2\chi)^{2}+4q_{0}^{2}}](k-\chi)^{\frac{5}{2}}+\mathcal{O}((k-\chi)^{\frac{7}{2}})\right],\qquad k\rightarrow\chi,\label{92}
\end{gather}
where
\begin{align}\label{93}
\omega(\chi)=2\left(\int_{\frac{iq_{0}}{2}-\frac{\beta}{2}}^{\chi}
+\int_{\frac{-iq_{0}}{2}-\frac{\beta}{2}}^{\chi}\right)\frac{y^{3}+h_{2}y^{2}+h_{1}y+h_{0}}{z(y)}dy.
\end{align}
We can find that the sign signature of Im$\omega(k)$ requires three branches of
Im$(\omega(k))=0$ emanating from point $\chi$, and through a straightforward calculation, it is not hard to present that
\begin{align}\label{94}
\int_{\frac{-iq_{0}}{2}-\frac{\beta}{2}}^{\frac{iq_{0}}{2}-\frac{\beta}{2}}
\frac{y^{3}+h_{2}y^{2}+h_{1}y+h_{0}}{z(y)}dy=\int_{\frac{-iq_{0}}{2}-\frac{\beta}{2}}^{\frac{iq_{0}}{2}-\frac{\beta}{2}}
\frac{\sqrt{(y-\chi_{1})^{2}+\chi_{2}^{2}}}{\sqrt{[(y+\frac{\beta}{2})^{2}+q_{0}^{2}]}}(y-k_{1}^{0})dy=0,
\end{align}
which indicates that the point $k_{1}^{0}$ is uniquely expressed. In Fig. 7, according to the sign signature of Im$(\omega)(k)$, the jump matrices $J_{3}^{(4)}, J_{5}^{(4)}$ and $J_{6}^{(4)}$ shall be bounded in the associated branch cuts.

Now, we consider the jump conditions of function $\omega(k)$, which satisfies
\begin{align}\label{95}
\omega_{+}(k)+\omega_{-}(k)=0,\qquad k\in \eta,\notag\\
\omega_{+}(k)+\omega_{-}(k)=\Lambda,\qquad k\in \varpi,
\end{align}
of which real constant $\Lambda$ can be given as
\begin{align}\label{96}
\Lambda=4\left(\int_{\frac{iq_{0}}{2}-\frac{\beta}{2}}^{\chi}
+\int_{\frac{-iq_{0}}{2}-\frac{\beta}{2}}^{\chi^{\ast}}\right)\frac{(y-k_{1}^{0})(y-\chi)(y-\chi^{\ast})}{z(y)}dy.
\end{align}
Further,
the following normalization condition of $\omega$ can be satisfied
\begin{align}\label{97}
\omega(k)=2k^{2}+(\xi+\beta)k+\omega_{0}+\mathcal{O}(\frac{1}{k}), \quad k\rightarrow\infty,
\end{align}
where $\omega_{0}$ is given in Eq.\eqref{87}.

In addition, the  function $\theta(k)$ has following  large-$k$ asymptotic
\begin{align}\label{98}
\theta(k)=2k^{2}+(\xi+\beta)k+\frac{1}{2}\beta\xi+q_{0}^{2}+\mathcal{O}(\frac{1}{k}), \quad k\rightarrow\infty.
\end{align}
Hence, from definition of function $\omega(k)$ in Eq. \eqref{71}, it is not hard to obtain
\begin{align}\label{99}
G(\infty)=\omega_{0}-\frac{1}{2}\beta\xi-q_{0}^{2}, \quad k\rightarrow\infty.
\end{align}

From Eq.\eqref{69}, we can derive the Riemann-Hilbert problem for $m^{(4)}$,
whose normalization condition is  $m^{(4)}\rightarrow e^{-iG(\infty)t\sigma_{3}}$ as $k\rightarrow\infty$. The contour $\widehat{\Sigma}^{(4)}$ is shown in Fig. 7 and the jump matrices $J^{(4)}$ are obtained as
\begin{gather}
J_{1}^{(4)}=\left(\begin{array}{cc}
   1  &  \delta^{2}\frac{\gamma^{\ast}e^{2i\omega t}}{1+\gamma\gamma^{\ast}}\\
    0&  1\\
\end{array}\right),\ J_{2}^{(4)}=\left(\begin{array}{cc}
   1  &  0\\
    \delta^{-2}\frac{\gamma e^{-2i\omega t}}{1+\gamma\gamma^{\ast}}&  1\\
\end{array}\right),
\notag\\
J_{3}^{(4)}=\left(\begin{array}{cc}
   1  &   0\\
  \delta^{-2}\gamma e^{-2i\omega t}  &  1\\
\end{array}\right),\ J_{4}^{(4)}=\left(\begin{array}{cc}
   1  &   \delta^{2}\gamma^{\ast} e^{2i\omega t}\\
    0 &  1\\
\end{array}\right),\notag\\
\ J_{\eta}^{(4)}=\left(\begin{array}{cc}
   0  &   \delta^{2}\frac{q_{-}}{iq_{0}}\\
  \delta^{-2}\frac{q^{\ast}_{-}}{iq_{0}} &  0\\
\end{array}\right),\notag\\
J_{5}^{(4)}=\left(\begin{array}{cc}
   1  &  \delta^{2}\gamma^{-1}e^{2i\omega t}\\
   0 &  1\\
\end{array}\right),\quad J_{6}^{(4)}=\left(\begin{array}{cc}
   1  &  0\\
   \delta^{-2}(\gamma^{\ast})^{-1}e^{-2i\omega t} &  1\\
\end{array}\right),\notag\\
J_{7}^{(4)}=\left(\begin{array}{cc}
   0  &  -\delta^{2}\gamma^{-1}e^{i\Lambda t}\\
  \delta^{-2}\gamma e^{-i\Lambda t} &  0\\
\end{array}\right),\notag\\ J_{8}^{(4)}=\left(\begin{array}{cc}
   0  &  \delta^{2}\gamma^{\ast}e^{i\Lambda t}\\
   -\delta^{-2}(\gamma^{\ast})^{-1}e^{-i\Lambda t} &  0\\
\end{array}\right). \label{100}
\end{gather}
\subsection{Further deformation}
Using $g$-function mechanism, the variable $k$ from the jump matrices  $J_{\eta}^{(4)}, J_{7}^{(4)}, J_{8}^{(4)}$ can be  eliminated. Therefore, we take following transformation
\begin{align}\label{101}
m^{(5)}=m^{(4)}e^{i\tilde{g}(k)\sigma_{3}},
\end{align}
of which the function $\tilde{g}(k)$ is analytic in $\mathbb{C}\setminus (\eta\cup\varpi)$, and it admits
\begin{align}\label{102}
\tilde{g}_{+}(k)+\tilde{g}_{-}(k)=\tilde{G}(k)=\left\{
\begin{array}{lr}
-i\ln(\delta^{2})\ \mbox{on} \ k\in\eta,\\
\vartheta-i\ln(\frac{\delta^{2}}{\gamma})\ \mbox{on} \ k\in\varpi_{+},\\
\vartheta-i\ln(\delta^{2}\gamma^{\ast})\ \mbox{on} \ k\in\varpi_{-},
  \end{array}
\right.
\end{align}
where $\vartheta$ is a  real constant. The normalization condition $\tilde{g}(z)\rightarrow \tilde{g}(\infty)$ as $k\rightarrow\infty$, it follows
\begin{align}\label{103a}
\frac{1}{2\pi i}\int_{\eta\cup\varpi}\frac{\tilde{G}(k,\vartheta)}{z_{-}(\zeta)}d\zeta=0.
\end{align}
In Eq.\eqref{103a},  the constant $\vartheta$ is determined.
Then the $\tilde{g}(\infty)$ is a real constant, given by
\begin{align}\label{103}
\tilde{g}(\infty)=\frac{1}{2\pi i}\int_{\eta\cup\varpi}\frac{\tilde{G}(k,\vartheta)}{z_{-}(\zeta)}\zeta d\zeta.
\end{align}

Finally, we obtain the following RH problem about $m^{(5)}$
\begin{align}\label{104a}
\left\{
\begin{array}{lr}
m^{(5)}(x, t, k)\ \mbox{is analytic in} \ \mathbb{C}\setminus\widehat{\Sigma}^{(5)},\\
m^{(5)}_{+}(x, t, k)=m^{(5)}_{-}(x, t, k)J^{(5)}(x, t, k), \qquad k\in\widehat{\Sigma}^{(5)},\\
m^{(5)}(x, t, k)\rightarrow e^{i(\tilde{g}(\infty)-G(\infty)t)\sigma_{3}},\qquad k\rightarrow \infty,
  \end{array}
\right.
\end{align}
where the contour $\widehat{\Sigma}^{(5)}=\widehat{\Sigma}^{(4)}$ is shown in Fig. 7 and the jump matrices $J^{(5)}$ are obtained as
\begin{gather}
J_{1}^{(5)}=\left(\begin{array}{cc}
   1  &  \delta^{2}\frac{\gamma^{\ast}e^{2i(\omega t-\tilde{g})}}{1+\gamma\gamma^{\ast}}\\
    0&  1\\
\end{array}\right),\ J_{2}^{(5)}=\left(\begin{array}{cc}
   1  &  0\\
    \delta^{-2}\frac{\gamma e^{-2i(\omega t--\tilde{g})}}{1+\gamma\gamma^{\ast}}&  1\\
\end{array}\right),
\notag\\
J_{3}^{(5)}=\left(\begin{array}{cc}
   1  &   0\\
  \delta^{-2}\gamma e^{-2i(\omega t-\tilde{g})}  &  1\\
\end{array}\right),\ J_{4}^{(5)}=\left(\begin{array}{cc}
   1  &   \delta^{2}\gamma^{\ast} e^{2i(\omega t-\tilde{g})}\\
    0 &  1\\
\end{array}\right),\notag\\
\ J_{\eta}^{(5)}=\left(\begin{array}{cc}
   0  &   \frac{q_{-}}{iq_{0}}\\
  \frac{q^{\ast}_{-}}{iq_{0}} &  0\\
\end{array}\right),\notag\\
J_{5}^{(5)}=\left(\begin{array}{cc}
   1  &  \delta^{2}\gamma^{-1}e^{2i(\omega t-\tilde{g})}\\
   0 &  1\\
\end{array}\right),\quad J_{6}^{(5)}=\left(\begin{array}{cc}
   1  &  0\\
   \delta^{-2}(\gamma^{\ast})^{-1}e^{-2i(\omega t-\tilde{g})} &  1\\
\end{array}\right),\notag\\
J_{7}^{(5)}=\left(\begin{array}{cc}
   0  &  -e^{i(\Lambda t-\vartheta)}\\
   e^{-i(\Lambda t-\vartheta)} &  0\\
\end{array}\right),\notag\\ J_{8}^{(5)}=\left(\begin{array}{cc}
   0  &  e^{i(\Lambda t-\vartheta)}\\
  -e^{-i(\Lambda t-\vartheta)} &  0\\
\end{array}\right). \label{104}
\end{gather}
\subsection{Model problem and the results}
From Fig. 7, it is not hard to find the jump matrices $J_{i}^{(5)}(i = 1, 2, 3, 4, 5, 6)$ is decaying exponentially to the identity away from the points $k_{1}^{0}$, $\chi$ and $\chi^{\ast}$ as $t\rightarrow\infty$. We know that the leading term of the solution is  determined by the model problem  which can be written as
\begin{align}\label{105}
\left\{
\begin{array}{lr}
m^{mod}(x, t, k)\ \mbox{is analytic in} \ \mathbb{C}\setminus (\eta\cup\varpi_{+}\cup(-\varpi_{-})),\\
m^{mod}_{+}(x, t, k)=m^{mod}_{-}(x, t, k)J^{mod}(x, t, k), \qquad k\in\eta\cup\varpi_{+}\cup(-\varpi_{-}),\\
m^{mod}(x, t, k)\rightarrow e^{i(\tilde{g}(\infty)-G(\infty)t)\sigma_{3}},\qquad k\rightarrow \infty,
  \end{array}
\right.
\end{align}
where
\begin{gather}
J_{\eta}^{mod}=J_{\eta}^{(5)}=\left(\begin{array}{cc}
   0  &   \frac{q_{-}}{iq_{0}}\\
  \frac{q^{\ast}_{-}}{iq_{0}} &  0\\
\end{array}\right),\notag\\
J_{\varpi_{+}\cup(-\varpi_{-})}^{mod}=\left(\begin{array}{cc}
   0  &  -e^{i(\Lambda t-\vartheta)}\\
   e^{-i(\Lambda t-\vartheta)} &  0\\
\end{array}\right), \label{106}
\end{gather}
and $-\varpi_{-}$ defines the opposite direction of cut $\varpi_{-}$.

For large $k$, introducing the factorization $m^{(5)}=m^{(err)}m^{(mod)}$ and using the Eq.\eqref{34}, we can obtain the solution $q(x,t)$ for the Eq.\eqref{3.1} in what follows
\begin{align}\label{107}
q(x,t)=-2i\left(m_{1}^{mod}(x,t)+m_{1}^{err}(x,t)\right)_{12}e^{i[\tilde{g}(\infty)-\beta x+(\alpha-2q_{0}^{2}-G(\infty))t]},
\end{align}
of which $\mid m_{1}^{err}\mid=\mathcal{O}(t^{-\frac{1}{2}})$.

Employing the elliptic theta functions, we can solve the model RH problem \eqref{105}. We first
consider the Abelian differential
\begin{align}\label{108}
d\vartheta=\frac{\vartheta_{0}}{z(k)}dk,\qquad \vartheta_{0}=\left(\oint_{L_{1}}\frac{1}{\vartheta(k)}dk\right)^{-1},
\end{align}
which can be normalized by taking $\oint_{L_{1}}d\vartheta=1$. As well as, the Abelian differential \eqref{108} has following Riemann period $\tau$
\begin{align}\label{109}
\tau=\oint_{L_{2}}d\vartheta.
\end{align}
It can be indicated from the results in \cite{Farkas} that $\tau$ is purely imaginary with $i\tau<0$. Considering the Abelian map
\begin{align}\label{110}
V(k)=\int_{iq_{0}-\frac{\beta}{2}}^{k}d\vartheta,
\end{align}
then we can give out the following relations
\begin{align}\label{111}
&V_{+}(k)+V_{-}(k)=n-\tau, \quad n\in\mathbb{Z},\quad k\in\varpi_{+}\cup(-\varpi_{-}),\notag\\
&V_{+}(k)+V_{-}(k)=n, \quad n\in\mathbb{Z},\quad k\in\eta,
\end{align}
and show a new function $r(k)$ as
\begin{align}\label{112}
r(k)=\left(\frac{(k-\chi)(k-iq_{0}+\frac{\beta}{2})}{(k-\chi^{\ast})(k+iq_{0}+\frac{\beta}{2})}\right)^{\frac{1}{4}},
\end{align}
which admits the same jump discontinuity across $\eta$ and $\varpi_{+}\cup(-\varpi_{-})$, as well as
$r_{+}(k)=ir_{-}(k)$. The function $r(k)$ has the following large-$k$ asymptotic
\begin{align}\label{113}
&r(k)=1-\frac{i(\chi_{2}+q_{0})}{2k}+O\left(\frac{1}{k^{2}}\right),\quad k\rightarrow\infty,\notag\\
&r(k)-r^{-1}(k)=-i(\chi_{2}+q_{0})+O\left(\frac{1}{k^{2}}\right),\quad k\rightarrow\infty.
\end{align}
As $i\tau<0$,  the theta function can be defined into
\begin{align}\label{114}
\Theta(k)=\sum_{\varrho\in\mathbb{Z}}e^{2\pi i\varrho k+\pi i\tau\varrho^{2}},
\end{align}
which yields
\begin{align}\label{115}
\Theta(k+n)=\Theta(k),\quad \Theta(k+n\tau)=e^{-(2\pi in k+\pi i\tau n^{2})}\Theta(k),\quad  n\in\mathbb{Z}.
\end{align}
Finally, the entries of the $2\times 2$ matrix-valued function $M(k)=M(x, t, k)$ are derived in what follows
\begin{align}\label{116}
M_{11}(k)=\frac{1}{2}[r(k)+r^{-1}(k)]\frac{\Theta\left(\frac{\Lambda t+\vartheta+i\ln\left(\frac{q_{-}^{\ast}}{iq_{0}}\right)}{2\pi}+V(k)+C\right)}
{\sqrt{\frac{iq_{0}}{q_{-}^{\ast}}}\Theta\left(V(k)+C\right)},\notag\\
M_{12}(k)=\frac{i}{2}[r(k)-r^{-1}(k)]\frac{\Theta\left(\frac{\Lambda t+\vartheta+i\ln\left(\frac{q_{-}^{\ast}}{iq_{0}}\right)}{2\pi}-V(k)+C\right)}
{\sqrt{\frac{q_{-}^{\ast}}{iq_{0}}}\Theta\left(-V(k)+C\right)},\notag\\
M_{21}(k)=\frac{-i}{2}[r(k)-r^{-1}(k)]\frac{\Theta\left(\frac{\Lambda t+\vartheta+i\ln\left(\frac{q_{-}^{\ast}}{iq_{0}}\right)}{2\pi}+V(k)-C\right)}
{\sqrt{\frac{iq_{0}}{q_{-}^{\ast}}}\Theta\left(V(k)-C\right)},\notag\\
M_{22}(k)=\frac{1}{2}[r(k)+r^{-1}(k)]\frac{\Theta\left(\frac{\Lambda t+\vartheta+i\ln\left(\frac{q_{-}^{\ast}}{iq_{0}}\right)}{2\pi}-V(k)-C\right)}
{\sqrt{\frac{q_{-}^{\ast}}{iq_{0}}}\Theta\left(-V(k)-C\right)},
\end{align}
of which
\begin{align}\label{117a}
C=V(k_{\ast})+\frac{1}{2}(1+\tau),\qquad k_{\ast}=\frac{2q_{0}\chi_{1}-\beta\chi_{2}}{2(q_{0}+\chi_{2})}.
\end{align}
Then the solution of the model RH problem \eqref{105} is given by
\begin{align}\label{117}
m^{mod}(x,t,k)=M^{-1}(x, t, \infty)M(x, t, k)e^{i(\tilde{g}(\infty)-G(\infty)t)\sigma_{3}},
\end{align}
then we have
\begin{align}\label{118}
(m^{mod})_{12}=\frac{iq_{0}}{2q_{-}^{\ast}}(\chi_{2}+q_{0})\frac{\Theta\left(\frac{\Lambda t+\vartheta+i\ln\left(\frac{q_{-}^{\ast}}{iq_{0}}\right)}{2\pi}-V(\infty)+C\right)
\Theta(V(\infty)+C)}{\Theta\left(\frac{\Lambda t+\vartheta+i\ln\left(\frac{q_{-}^{\ast}}{iq_{0}}\right)}{2\pi}+V(\infty)+C\right)\Theta(-V(\infty)+C)}e^{i(\tilde{g}(\infty)
-G(\infty)t)}.
\end{align}
Based on Eq.\eqref{107}, we have
\begin{gather}
q(x,t)=\frac{q_{0}(\chi_{2}+q_{0})}{q_{-}^{\ast}}\frac{\Theta\left(\frac{\Lambda t+\vartheta+i\ln\left(\frac{q_{-}^{\ast}}{iq_{0}}\right)}{2\pi}-V(\infty)+C\right)
\Theta(V(\infty)+C)}{\Theta\left(\frac{\Lambda t+\vartheta+i\ln\left(\frac{q_{-}^{\ast}}{iq_{0}}\right)}{2\pi}+V(\infty)+C\right)\Theta(-V(\infty)+C)}\notag\\
e^{i[2\tilde{g}(\infty)-\beta x+(\alpha-2q_{0}^{2}-2G(\infty))t]}+\mathcal{O}(t^{-\frac{1}{2}}),\label{119}
\end{gather}
where $V(\infty)=\int_{iq_{0}-\frac{\beta}{2}}^{\infty}d\vartheta$, and $\chi_{2}, \Lambda, G(\infty), \vartheta, \tilde{g}(\infty), C$ are given by Eqs. \eqref{84}, \eqref{96}, \eqref{99}, \eqref{103a}, \eqref{103}, \eqref{117a}, respectively.

\section{Plane wave region II}
When $\xi<\xi_{1}$, there are two real roots $k_{1}^{\pm}$ on the real axis, and they both distribute on the right side of the $\eta$ cut. In order to discuss the long-time asymptotics of solution for the Eq.\eqref{3.1} with boundary conditions \eqref{4}, we first analyse the original RHP $m^{(0)}$ and deform the contour $\Sigma^{(0)}$.
\subsection{A transformation between $m^{(0)}$ and $\tilde{m}^{(0)}$}
Firstly,   the RH problem \eqref{32} needs to be rescaled into
\begin{align}\label{40a}
\tilde{m}^{(0)}=\left\{
\begin{array}{lr}
m^{(0)}H^{\ast}(k),\qquad \ k\in \mathbb{C}_{+}\setminus\eta_{+},\\
m^{(0)}H^{-1}(k), \qquad \ k\in \mathbb{C}_{-}\setminus\eta_{-},
  \end{array}
\right.
\end{align}
where
\begin{align}\label{41a}
H(k)=
\left(\begin{array}{cc}
   a(k)  &  0\\
   0 &  a(k)^{-1}\\
\end{array}\right).
\end{align}
Then the matrix-value function $\tilde{m}^{(0)}(x, t, k)$ is the solution to the following Riemann-Hilbert problem:
\begin{align}\label{42a}
\left\{
\begin{array}{lr}
\tilde{m}^{(0)}(x, t, k)\ \mbox{is analytic in} \ \mathbb{C}\setminus\tilde{\Sigma}^{(0)},\\
\tilde{m}^{(0)}_{+}(x, t, k)=\tilde{m}^{(0)}_{-}(x, t, k)\tilde{J}^{(0)}(x, t, k), \qquad k\in\tilde{\Sigma}^{(0)}=\Sigma^{(0)},\\
\tilde{m}^{(0)}(x, t, k)\rightarrow I,\qquad k\rightarrow \infty,
  \end{array}
\right.
\end{align}
of which the jump matrix reads
\begin{gather}
\tilde{J}_{1}^{(0)}=HJ_{1}^{(0)}H^{\ast},\quad
\tilde{J}_{2}^{(0)}=(H_{-}^{\ast})^{-1}J_{2}^{(0)}H_{+}^{\ast},\quad
\tilde{J}_{3}^{(0)}=H_{-}J_{3}^{(0)}H_{+}^{-1}.\label{43a}
\end{gather}
\subsection{A transformation between $\tilde{m}^{(1)}$ and $\tilde{m}^{(2)}$}
Analogously, we can factor  the jump matrices $\tilde{J}_{1}^{(0)}, \tilde{J}_{2}^{(0)}, \tilde{J}_{3}^{(0)}$ into
\begin{align}
&\tilde{J}_{1}^{(0)}=\tilde{J}_{2}^{(1)}\tilde{J}_{0}^{(1)}\tilde{J}_{1}^{(1)},\qquad \quad \ \mbox{on} \qquad (k_{1}^{+},\infty),\notag\\
&\tilde{J}_{1}^{(0)}=\tilde{J}_{4}^{(1)}\tilde{J}_{3}^{(1)},\qquad\qquad \quad \mbox{on} \qquad ( -\infty,k_{1}^{+}),\notag\\
&\tilde{J}_{2}^{(0)}=(\tilde{J}_{3-}^{(1)})^{-1}\tilde{J}_{\eta}^{(1)}\tilde{J}_{3+}^{(1)},\qquad \mbox{on} \qquad \eta_{+} \qquad \mbox{cut},\notag\\
&\tilde{J}_{3}^{(0)}=\tilde{J}_{4-}^{(1)}\tilde{J}_{\eta}^{(1)}(\tilde{J}_{4+}^{(1)})^{-1},\qquad \mbox{on} \qquad \eta_{-} \qquad \mbox{cut},\notag
\end{align}
where
\begin{gather}
\tilde{J}_{0}^{(1)}=\left(\begin{array}{cc}
  \frac{1}{1+\rho\rho^{\ast}}  &  0\\
   0 &  1+\rho\rho^{\ast}\\
\end{array}\right),\
\tilde{J}_{1}^{(1)}=\left(\begin{array}{cc}
   \Delta^{-\frac{1}{2}}  &  0\\
    \frac{\Delta^{-\frac{1}{2}}\rho e^{-2i\theta t}}{1+\rho\rho^{\ast}}&  \Delta^{\frac{1}{2}}\\
\end{array}\right),\ \tilde{J}_{2}^{(1)}=\left(\begin{array}{cc}
   \Delta^{-\frac{1}{2}}  &  \frac{\Delta^{-\frac{1}{2}}\rho^{\ast}e^{2i\theta t}}{1+\rho\rho^{\ast}}\\
   0 &  \Delta^{\frac{1}{2}}\\
\end{array}\right),
\notag\\
\tilde{J}_{3}^{(1)}=\left(\begin{array}{cc}
   \Delta^{-\frac{1}{2}}  &   \Delta^{\frac{1}{2}}\rho^{\ast} e^{2i\theta t}\\
   0 &  \Delta^{\frac{1}{2}}\\
\end{array}\right),\ \tilde{J}_{4}^{(1)}=\left(\begin{array}{cc}
   \Delta^{-\frac{1}{2}}  &  0 \\
    \Delta^{\frac{1}{2}}\rho e^{-2i\theta t} &  \Delta^{\frac{1}{2}}\\
\end{array}\right),\ \tilde{J}_{\eta}^{(1)}=\left(\begin{array}{cc}
   0  &   \frac{q_{+}}{iq_{0}}\\
  \frac{q^{\ast}_{+}}{iq_{0}} &  0\\
\end{array}\right), \label{40b}
\end{gather}
and we have defined $\rho(k)=-\frac{b(k)}{a(k)}$.
Taking
\begin{align}\label{41b}
\tilde{m}^{(1)}=\tilde{m}^{(0)}\tilde{B}(k),
\end{align}
where
\begin{align}\label{42b}
\tilde{B}(k)=\left\{
\begin{array}{lr}
(\tilde{J}_{1}^{(1)})^{-1}\ \mbox{on} \ k\in\tilde{\Omega}_{1},\\
\tilde{J}_{2}^{(1)}\ \mbox{on} \ k\in\tilde{\Omega}_{2},\\
(\tilde{J}_{3}^{(1)})^{-1}\ \mbox{on} \ k\in\tilde{\Omega}_{3}\cup\tilde{\Omega}_{5},\\
\tilde{J}_{4}^{(1)}\ \mbox{on} \ k\in\tilde{\Omega}_{4}\cup\tilde{\Omega}_{6},\\
I \ \mbox{on} \ k\in \mbox{others},
  \end{array}
\right.
\end{align}
we derive a new contour $\tilde{\Sigma}^{(1)}$ of the new matrix-value function $\tilde{m}^{(1)}$  from the contour $\tilde{\Sigma}^{(0)}$  in Fig. 8. The jump matrix of the new matrix-value function $\tilde{m}^{(1)}$ is $\tilde{J}^{(1)}$, given in \eqref{40b} (see Fig. 9).\\

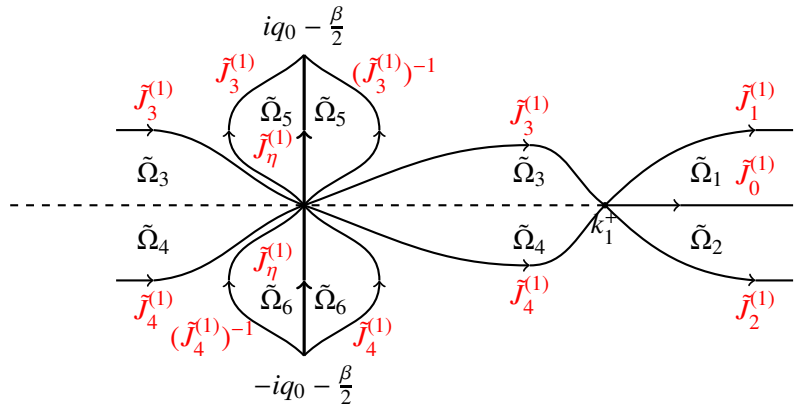
\centerline{\begin{tikzpicture}
\draw[-][thick](6.5,1)--(6,1);
\draw [<-,thick] (6,1) to [out=-175,in=45] (4,0);
\draw [-,thick] (4,0) to [out=-135,in=0] (3,-0.8);
\draw [<-,thick] (3,-0.8) to [out=180,in=-20] (0,0);
\draw [-,thick] (0,0) to [out=160,in=-5] (-2,1);
\draw[<-][thick](-2,1)--(-2.5,1);
\draw[-][thick](6.5,-1)--(6,-1);
\draw [<-,thick] (6,-1) to [out=175,in=-45] (4,0);
\draw [-,thick] (4,0) to [out=145,in=0] (3,0.8);
\draw [<-,thick] (3,0.8) to [out=180,in=20] (0,0);
\draw [-,thick] (0,0) to [out=-160,in=5] (-2,-1);
\draw[<-][thick](-2,-1)--(-2.5,-1);
\draw[-][thick](6.5,0)--(5,0);
\draw[<-][thick](5,0)--(4,0);
\draw[-][dashed,thick](4,0)--(-4,0);
\draw[->][very thick](0,0)--(0,1);
\draw[-][very thick](0,1)--(0,2);
\draw[-][very thick](0,0)--(0,-1);
\draw[<-][very thick](0,-1)--(0,-2);
\draw [->,thick] (0,0) to [out=50,in=-90] (1,1);
\draw [-,thick]  (1,1) to [out=90,in=-45] (0,2);
\draw [-,thick] (0,0) to [out=-50,in=90] (1,-1);
\draw [<-,thick]  (1,-1) to [out=-90,in=45] (0,-2);
\draw [->,thick] (0,0) to [out=130,in=-90] (-1,1);
\draw [-,thick]  (-1,1) to [out=90,in=-135] (0,2);
\draw [-,thick] (0,0) to [out=-130,in=90] (-1,-1);
\draw [<-,thick]  (-1,-1) to [out=-90,in=135] (0,-2);
\draw[fill] (5,0.45) node[right]{$\tilde{\Omega}_{1}$};
\draw[fill] (5,-0.45) node[right]{$\tilde{\Omega}_{2}$};
\draw[fill] (3,0.45) node{$\tilde{\Omega}_{3}$};
\draw[fill] (3,-0.45) node{$\tilde{\Omega}_{4}$};
\draw[fill] (-2,0.45) node{$\tilde{\Omega}_{3}$};
\draw[fill] (-2,-0.45) node{$\tilde{\Omega}_{4}$};
\draw[fill] (0,1.25) node[right]{$\tilde{\Omega}_{5}$};
\draw[fill] (0,-1.25) node[right]{$\tilde{\Omega}_{6}$};
\draw[fill] (0,1.25) node[left]{$\tilde{\Omega}_{5}$};
\draw[fill] (0,-1.25) node[left]{$\tilde{\Omega}_{6}$};
\draw[fill] (4,-0.3) node{$k_{1}^{+}$};
\draw[fill] (4,0) circle [radius=0.035];
\draw[fill] (0,-2) node[below]{$-iq_{0}-\frac{\beta}{2}$};
\draw[fill] (0,2) node[above]{$iq_{0}-\frac{\beta}{2}$};
\draw[fill] (6,1) node[above]{$\textcolor[rgb]{1.00,0.00,0.00}{\tilde{J}_{1}^{(1)}}$};
\draw[fill] (6,-1) node[below]{$\textcolor[rgb]{1.00,0.00,0.00}{\tilde{J}_{2}^{(1)}}$};
\draw[fill] (6,0) node[above]{$\textcolor[rgb]{1.00,0.00,0.00}{\tilde{J}_{0}^{(1)}}$};
\draw[fill] (3,0.8) node[above]{$\textcolor[rgb]{1.00,0.00,0.00}{\tilde{J}_{3}^{(1)}}$};
\draw[fill] (3,-0.8) node[below]{$\textcolor[rgb]{1.00,0.00,0.00}{\tilde{J}_{4}^{(1)}}$};
\draw[fill] (-2,1) node[above]{$\textcolor[rgb]{1.00,0.00,0.00}{\tilde{J}_{3}^{(1)}}$};
\draw[fill] (-2,-1) node[below]{$\textcolor[rgb]{1.00,0.00,0.00}{\tilde{J}_{4}^{(1)}}$};
\draw[fill] (0.5,1.75) node[right]{$\textcolor[rgb]{1.00,0.00,0.00}{(\tilde{J}_{3}^{(1)})^{-1}}$};
\draw[fill] (0.5,-1.75) node[right]{$\textcolor[rgb]{1.00,0.00,0.00}{\tilde{J}_{4}^{(1)}}$};
\draw[fill] (-0.5,1.75) node[left]{$\textcolor[rgb]{1.00,0.00,0.00}{\tilde{J}_{3}^{(1)}}$};
\draw[fill] (-0.5,-1.75) node[left]{$\textcolor[rgb]{1.00,0.00,0.00}{(\tilde{J}_{4}^{(1)})^{-1}}$};
\draw[fill] (0,0.75) node[left]{$\textcolor[rgb]{1.00,0.00,0.00}{\tilde{J}_{\eta}^{(1)}}$};
\draw[fill] (0,-0.75) node[left]{$\textcolor[rgb]{1.00,0.00,0.00}{\tilde{J}_{\eta}^{(1)}}$};
\end{tikzpicture}}
\centerline{\noindent {\small \textbf{Figure 8.} (Color online) The  contour  $\tilde{\Sigma}^{(0)}$.}}

\centerline{\begin{tikzpicture}
\draw[-][thick](6.5,1)--(6,1);
\draw [<-,thick] (6,1) to [out=-175,in=45] (4,0);
\draw [-,thick] (4,0) to [out=-45,in=0] (2,1.2);
\draw[<-][thick](2,1.2)--(0,1.2);
\draw [-,thick] (0,1.2) to [out=-180,in=15] (-1,0.3);
\draw [<-,thick] (-1,0.3) to [out=-165,in=0] (-2,0.1);
\draw[-][thick](6.5,-1)--(6,-1);
\draw [<-,thick] (6,-1) to [out=175,in=-45] (4,0);
\draw [-,thick] (4,0) to [out=-135,in=0] (2,-1.2);
\draw[<-][thick](2,-1.2)--(0,-1.2);
\draw [-,thick] (0,-1.2) to [out=180,in=-15] (-1,-0.3);
\draw [<-,thick] (-1,-0.3) to [out=165,in=0] (-2,-0.1);
\draw[-][thick](6.5,0)--(5,0);
\draw[<-][thick](5,0)--(4,0);
\draw[->][very thick](2,0)--(2,0.4);
\draw[-][very thick](2,0.4)--(2,0.8);
\draw[-][very thick](2,0)--(2,-0.4);
\draw[-][very thick](2,-0.4)--(2,-0.8);
\draw[fill] (4,-0.3) node{$k_{1}^{+}$};
\draw[fill] (4,0) circle [radius=0.035];
\draw[fill] (2,-0.8) node[left]{$-iq_{0}-\frac{\beta}{2}$};
\draw[fill] (2,0.8) node[left]{$iq_{0}-\frac{\beta}{2}$};
\draw[fill] (6,1) node[above]{$\textcolor[rgb]{1.00,0.00,0.00}{\tilde{J}_{1}^{(1)}}$};
\draw[fill] (6,-1) node[below]{$\textcolor[rgb]{1.00,0.00,0.00}{\tilde{J}_{2}^{(1)}}$};
\draw[fill] (6,0) node[above]{$\textcolor[rgb]{1.00,0.00,0.00}{\tilde{J}_{0}^{(1)}}$};
\draw[fill] (2,0.4) node[right]{$\textcolor[rgb]{1.00,0.00,0.00}{\tilde{J}_{\eta}^{(1)}}$};
\draw[fill] (2,1.2) node[above]{$\textcolor[rgb]{1.00,0.00,0.00}{\tilde{J}_{3}^{(1)}}$};
\draw[fill] (2,-1.2) node[below]{$\textcolor[rgb]{1.00,0.00,0.00}{\tilde{J}_{4}^{(1)}}$};
\end{tikzpicture}}
\centerline{\noindent {\small \textbf{Figure 9.} (Color online) The  contour  $\tilde{\Sigma}^{(1)}$.}}
\subsection{Eliminating the jump across the cut $(k_{1}^{+},\infty)$}
To eliminating the jump across the cut $(k_{1}^{+},\infty)$, we first show the function $\delta(k)$, that  is analytic in $\mathbb{C}\setminus (k_{1}^{+},\infty)$ and meets the jump condition
\begin{align}\label{43b}
\delta_{+}(k)=\delta_{-}(k)\frac{1}{1+\rho(k)\rho^{\ast}(k)},\qquad k\in (k_{1}^{+},\infty),
\end{align}
and the normalization condition
\begin{align}\label{44b}
\delta(k)=1+\mathcal{O}(\frac{1}{k}),\qquad k\rightarrow\infty.
\end{align}
The solution $\delta(k)$ of the RH problem \eqref{43b} is
\begin{align}\label{45b}
\delta(k)=\exp\big\{\frac{1}{2\pi i}\int_{k_{1}^{+}}^{\infty}\frac{\ln\frac{1}{1+\rho(y)\rho^{\ast}(y)}}{y-k}dy\big\}.
\end{align}
Setting
\begin{align}\label{46b}
\tilde{m}^{(2)}=\tilde{m}^{(1)}\delta^{-\sigma_{3}},
\end{align}
we can obtain $\tilde{J}^{(2)}=\delta_{-}^{\sigma_{3}}\tilde{J}^{(1)}\delta_{+}^{-\sigma_{3}}$, and
\begin{gather}
\tilde{J}_{1}^{(2)}=\left(\begin{array}{cc}
   \Delta^{-\frac{1}{2}}  &  0\\
    \delta^{-2}\frac{\Delta^{-\frac{1}{2}}\rho e^{-2i\theta t}}{1+\rho\rho^{\ast}}&  \Delta^{\frac{1}{2}}\\
\end{array}\right),\ \tilde{J}_{2}^{(2)}=\left(\begin{array}{cc}
   \Delta^{-\frac{1}{2}}  &  \delta^{2}\frac{\Delta^{-\frac{1}{2}}\rho^{\ast}e^{2i\theta t}}{1+\rho\rho^{\ast}}\\
   0 &  \Delta^{\frac{1}{2}}\\
\end{array}\right),
\notag\\
\tilde{J}_{3}^{(2)}=\left(\begin{array}{cc}
   \Delta^{-\frac{1}{2}}  &   \delta^{2}\Delta^{\frac{1}{2}}\rho^{\ast} e^{2i\theta t}\\
   0 &  \Delta^{\frac{1}{2}}\\
\end{array}\right),\ \tilde{J}_{4}^{(2)}=\left(\begin{array}{cc}
   \Delta^{-\frac{1}{2}}  &  0 \\
   \delta^{-2}\Delta^{\frac{1}{2}}\rho e^{-2i\theta t} &  \Delta^{\frac{1}{2}}\\
\end{array}\right),\ \tilde{J}_{\eta}^{(2)}=\left(\begin{array}{cc}
   0  &   \delta^{2}\frac{q_{+}}{iq_{0}}\\
  \delta^{-2}\frac{q^{\ast}_{+}}{iq_{0}} &  0\\
\end{array}\right). \label{47b}
\end{gather}
A new matrix-value function $\tilde{m}^{(2)}$ satisfies the following RH problem
\begin{align}\label{47.1b}
\left\{
\begin{array}{lr}
\tilde{m}^{(2)}(x, t, k)\ \mbox{is analytic in} \ \mathbb{C}\setminus\tilde{\Sigma}^{(2)},\\
\tilde{m}^{(2)}_{+}(x, t, k)=\tilde{m}^{(2)}_{-}(x, t, k)\tilde{J}^{(2)}(x, t, k), \qquad k\in\tilde{\Sigma}^{(2)},\\
\tilde{m}^{(2)}(x, t, k)\rightarrow I,\qquad k\rightarrow \infty,
  \end{array}
\right.
\end{align}
where  $\tilde{J}^{(2)}$ is given in Eq. \eqref{47b} and the contour $\tilde{\Sigma}^{(2)}$ is shown in Fig. 10.

\centerline{\begin{tikzpicture}
\draw[-][thick](6.5,1)--(6,1);
\draw [<-,thick] (6,1) to [out=-175,in=45] (4,0);
\draw [-,thick] (4,0) to [out=-45,in=0] (2,1.2);
\draw[<-][thick](2,1.2)--(0,1.2);
\draw [-,thick] (0,1.2) to [out=-180,in=15] (-1,0.3);
\draw [<-,thick] (-1,0.3) to [out=-165,in=0] (-2,0.1);
\draw[-][thick](6.5,-1)--(6,-1);
\draw [<-,thick] (6,-1) to [out=175,in=-45] (4,0);
\draw [-,thick] (4,0) to [out=-135,in=0] (2,-1.2);
\draw[<-][thick](2,-1.2)--(0,-1.2);
\draw [-,thick] (0,-1.2) to [out=180,in=-15] (-1,-0.3);
\draw [<-,thick] (-1,-0.3) to [out=165,in=0] (-2,-0.1);
\draw[->][very thick](2,0)--(2,0.4);
\draw[-][very thick](2,0.4)--(2,0.8);
\draw[-][very thick](2,0)--(2,-0.4);
\draw[-][very thick](2,-0.4)--(2,-0.8);
\draw[fill] (4,-0.3) node{$k_{1}^{+}$};
\draw[fill] (4,0) circle [radius=0.035];
\draw[fill] (2,-0.8) node[left]{$-iq_{0}-\frac{\beta}{2}$};
\draw[fill] (2,0.8) node[left]{$iq_{0}-\frac{\beta}{2}$};
\draw[fill] (6,1) node[above]{$\textcolor[rgb]{1.00,0.00,0.00}{\tilde{J}_{1}^{(2)}}$};
\draw[fill] (6,-1) node[below]{$\textcolor[rgb]{1.00,0.00,0.00}{\tilde{J}_{2}^{(2)}}$};
\draw[fill] (6,0) node[above]{$\textcolor[rgb]{1.00,0.00,0.00}{\tilde{J}_{0}^{(2)}}$};
\draw[fill] (2,0.4) node[right]{$\textcolor[rgb]{1.00,0.00,0.00}{\tilde{J}_{\eta}^{(2)}}$};
\draw[fill] (2,1.2) node[above]{$\textcolor[rgb]{1.00,0.00,0.00}{\tilde{J}_{3}^{(2)}}$};
\draw[fill] (2,-1.2) node[below]{$\textcolor[rgb]{1.00,0.00,0.00}{\tilde{J}_{4}^{(2)}}$};
\draw[fill] (5,0) node[right]{$\widehat{\tilde{\Omega}}_{4}$};
\draw[fill] (4,0.5) node[above]{$\widehat{\tilde{\Omega}}_{1}$};
\draw[fill] (4,-0.5) node[below]{$\widehat{\tilde{\Omega}}_{2}$};
\draw[fill] (1,0) node{$\widehat{\tilde{\Omega}}_{3}$};
\end{tikzpicture}}
\centerline{\noindent {\small \textbf{Figure 10.} (Color online) The  contour  $\tilde{\Sigma}^{(2)}$.}}

\subsection{Removing the term $\Delta(k)$}
In order to remove the term $\Delta(k)$, we need take a transformation, given by
\begin{align}\label{48b}
\tilde{m}^{(3)}=\tilde{m}^{(2)}\widehat{\tilde{B}}(k),
\end{align}
where
\begin{align}\label{49b}
\widehat{\tilde{B}}(k)=\left\{
\begin{array}{lr}
\Delta^{\frac{\sigma_{3}}{2}}\ \mbox{on} \ k\in\widehat{\tilde{\Omega}}_{1},\\
\Delta^{-\frac{\sigma_{3}}{2}}\ \mbox{on} \ k\in\widehat{\tilde{\Omega}}_{2},\\
I \ \mbox{on} \ k\in \widehat{\tilde{\Omega}}_{3}\cup\widehat{\tilde{\Omega}}_{4}.
  \end{array}
\right.
\end{align}
Therefore,
we obtain the following RH problem for $\tilde{\Sigma}^{(3)}$:
\begin{align}\label{49.1b}
\left\{
\begin{array}{lr}
\tilde{m}^{(3)}(x, t, k)\ \mbox{is analytic in} \ \mathbb{C}\setminus\tilde{\Sigma}^{(3)},\\
\tilde{m}^{(3)}_{+}(x, t, k)=\tilde{m}^{(3)}_{-}(x, t, k)\tilde{J}^{(3)}(x, t, k), \qquad k\in\tilde{\Sigma}^{(3)},\\
\tilde{m}^{(3)}(x, t, k)\rightarrow I,\qquad k\rightarrow \infty,
  \end{array}
\right.
\end{align}
where the contour $\tilde{\Sigma}^{(3)}=\tilde{\Sigma}^{(2)}$ is shown in Fig. 10, and $\tilde{J}^{(3)}$ reads
\begin{gather}
\tilde{J}_{1}^{(3)}=\left(\begin{array}{cc}
   1  &  0\\
    \delta^{-2}\frac{\rho e^{-2i\theta t}}{1+\rho\rho^{\ast}}&  1\\
\end{array}\right),\ \tilde{J}_{2}^{(3)}=\left(\begin{array}{cc}
   1  &  \delta^{2}\frac{\rho^{\ast}e^{2i\theta t}}{1+\rho\rho^{\ast}}\\
   0 &  1\\
\end{array}\right),
\notag\\
\tilde{J}_{3}^{(3)}=\left(\begin{array}{cc}
   1  &   \delta^{2}\rho^{\ast} e^{2i\theta t}\\
   0 &  1\\
\end{array}\right),\ \tilde{J}_{4}^{(3)}=\left(\begin{array}{cc}
   1  &  0 \\
   \delta^{-2}\rho e^{-2i\theta t} &  1\\
\end{array}\right),\ \tilde{J}_{\eta}^{(3)}=\tilde{J}_{\eta}^{(2)}. \label{50b}
\end{gather}
\subsection{The $g$-function and model problem}
In the same way,  a $g$-function is  introduced to make the following transformation
\begin{align}\label{51b}
\tilde{m}^{(4)}=\tilde{m}^{(3)}e^{ig(k)\sigma_{3}},
\end{align}
where $g(k)$ is analytic in $\mathbb{C}\setminus \eta$, and has
the following discontinuity condition
\begin{align}\label{52b}
\delta^{-2}(k)e^{i(g_{+}(k)+g_{-}(k))}=1,\qquad k\in \eta,
\end{align}
then the jump matrix $\tilde{J}_{\eta}^{(4)}$ across $\eta$ is
\begin{align}\label{53b}
\tilde{J}_{\eta}^{(4)}=\left(\begin{array}{cc}
   0  &   \frac{q_{+}}{iq_{0}}\\
  \frac{q^{\ast}_{+}}{iq_{0}} &  0\\
\end{array}\right),\qquad k\in \eta,
\end{align}
which  become a constant.
Besides, the other jump matrices in Eq.\eqref{50b} turn into
\begin{gather}
\tilde{J}_{1}^{(4)}=\left(\begin{array}{cc}
   1  &  0\\
    \delta^{-2}\frac{\rho e^{-2i(\theta t-g)}}{1+\rho\rho^{\ast}}&  1\\
\end{array}\right),\ \tilde{J}_{2}^{(4)}=\left(\begin{array}{cc}
   1  &  \delta^{2}\frac{\rho^{\ast}e^{2i(\theta t-g)}}{1+\rho\rho^{\ast}}\\
   0 &  1\\
\end{array}\right),
\notag\\
\tilde{J}_{3}^{(4)}=\left(\begin{array}{cc}
   1  &   \delta^{2}\rho^{\ast} e^{2i(\theta t-g)}\\
   0 &  1\\
\end{array}\right),\ \tilde{J}_{4}^{(4)}=\left(\begin{array}{cc}
   1  &  0 \\
   \delta^{-2}\rho e^{-2i(\theta t-g)} &  1\\
\end{array}\right). \label{54b}
\end{gather}
On the $\eta$ cut, one has $\lambda_{-}=-\lambda_{+}$. According to the Plemelj's formula, we have
\begin{align}\label{54.1b}
g(k)=\frac{\lambda(k)}{2\pi^{2}i}\int_{z\in \eta}\frac{1}{\lambda(z)(z-k)}\int_{k_{1}^{+}}^{\infty}\frac{\ln\frac{1}{1+\rho(y)\rho^{\ast}(y)}}{y-z}dydz,
\end{align}
Then, as $k\rightarrow\infty$, we get
\begin{align}\label{55b}
g(\infty)=-\frac{1}{2\pi^{2}i}\int_{z\in \eta}\frac{1}{\lambda(z)}\int_{k_{1}^{+}}^{\infty}\frac{\ln\frac{1}{1+\rho(y)\rho^{\ast}(y)}}{y-z}dydz,
\end{align}
which is only dependent of $q_{0}, \beta$ and $\xi$, whereas is independent of $k$.
The jump matrices $\tilde{J}_{i}^{(4)}, (i = 1, 2, 3, 4)$ decay exponentially to the identity away from the point $k_{1}^{+}$ as $t\rightarrow\infty$. Finally, one can write $\tilde{m}^{(4)}$ in the form
\begin{align}\label{56b}
\tilde{m}^{(4)}=\tilde{m}^{err}\tilde{m}^{mod},
\end{align}
where $\tilde{m}^{err}$ problem yields
\begin{align}\label{57b}
\tilde{m}^{err}=I+\mathcal{O}(t^{-\frac{1}{2}}).
\end{align}
Obviously, the model problem will dominate the long-time asymptotics
of the solution, and $\tilde{m}^{mod}$ solve the following RH problem:
\begin{align}\label{58b}
\left\{
\begin{array}{lr}
\tilde{m}^{mod}(x, t, k)\ \mbox{is analytic in} \ \mathbb{C}\setminus \eta,\\
\tilde{m}^{mod}_{+}(x, t, k)=\tilde{m}^{mod}_{-}(x, t, k)\tilde{J}^{mod}(x, t, k), \qquad k\in\eta,\\
\tilde{m}^{mod}(x, t, k)\rightarrow e^{ig(\infty)\sigma_{3}},\qquad k\rightarrow \infty,
  \end{array}
\right.
\end{align}
of which the jump matrix $\tilde{J}^{mod}(x, t, k)=\tilde{J}_{\eta}^{(4)}(x, t, k)$ given in \eqref{53b}, which has exact solution
\begin{align}\label{59b}
\tilde{m}^{mod}=e^{ig(\infty)\sigma_{3}}\left(\begin{array}{cc}
   \frac{1}{2}(\Upsilon+\Upsilon^{-1})  &   -\frac{q_{0}}{2q^{\ast}_{+}}(\Upsilon-\Upsilon^{-1})\\
  \frac{q_{0}}{2q_{+}}(\Upsilon-\Upsilon^{-1})^{\ast}  &  \frac{1}{2}(\Upsilon+\Upsilon^{-1})^{\ast}\\
\end{array}\right),
\end{align}
where
\begin{align}\label{60b}
\Upsilon(k)=\left(\frac{k-iq_{0}+\frac{\beta}{2}}{k+iq_{0}+\frac{\beta}{2}}\right)^{\frac{1}{4}}.
\end{align}
Then, based on the solution $q(x, t)$ in \eqref{34}, one has
\begin{align}\label{61b}
q(x,t)&=-2i\left(\tilde{m}_{1}^{(0)}(x,t)\right)_{12}e^{i[-\beta x+(\alpha-2q_{0}^{2})t]}\notag\\
&=-2i\left(\tilde{m}_{1}^{(3)}(x,t)\right)_{12}e^{i[-\beta x+(\alpha-2q_{0}^{2})t]}\notag\\
&=-2i\left(\tilde{m}_{1}^{mod}(x,t)e^{-ig(\infty)\sigma_{3}}\right)_{12}e^{i[-\beta x+(\alpha-2q_{0}^{2})t]}+\mathcal{O}(t^{-\frac{1}{2}}),
\end{align}
where $\tilde{m}_{1}^{mod}$ is shown by the explicit solution $\tilde{m}^{mod}$ in Eq. \eqref{59b}:
\begin{align}\label{62b}
\tilde{m}_{1}^{mod}&=\lim\limits_{k\rightarrow \infty}k(\tilde{m}^{mod}-e^{ig(\infty)\sigma_{3}})\notag\\
&=e^{ig(\infty)\sigma_{3}}\lim\limits_{k\rightarrow \infty}k\left(\begin{array}{cc}
   \frac{1}{2}(\Upsilon+\Upsilon^{-1})-1  &   -\frac{q_{0}}{2q^{\ast}_{+}}(\Upsilon-\Upsilon^{-1})\\
  \frac{q_{0}}{2q_{+}}(\Upsilon-\Upsilon^{-1})^{\ast}  &  \frac{1}{2}(\Upsilon+\Upsilon^{-1})^{\ast}-1\\
\end{array}\right).
\end{align}
Observing that
\begin{align}\label{63b}
\Upsilon-\Upsilon^{-1}=\frac{-iq_{0}}{k}+O(\frac{1}{k}),\qquad \mbox{as} \quad k\rightarrow\infty,
\end{align}
then we have
\begin{align}\label{64b}
\left(\tilde{m}_{1}^{mod}(x,t)e^{-ig(\infty)\sigma_{3}}\right)_{12}=-e^{2ig(\infty)}\lim\limits_{k\rightarrow \infty}k\frac{q_{0}}{2q^{\ast}_{+}}(\Upsilon-\Upsilon^{-1})=\frac{iq_{+}}{2}e^{2ig(\infty)}.
\end{align}
Hence, going back to Eq.\eqref{61b}, in the plane wave region II, we derive the long-time asymptotic behavior of the solution for the Eq.\eqref{3.1}
\begin{align}\label{65b}
q(x,t)=q_{+}e^{i[2g(\infty)-\beta x+(\alpha-2q_{0}^{2})t]}+\mathcal{O}(t^{-\frac{1}{2}}),
\end{align}
where $g(\infty)$ is given by Eq.\eqref{55b}, which is only dependent of $q_{0}, \beta$ and $\xi$ with $\xi<\xi_{1}$.

\section{Acknowledgements}
This work was supported by the Postgraduate Research and Practice of Educational Reform for Graduate students in CUMT under Grant No. 2019YJSJG046, the Natural Science Foundation of Jiangsu Province under Grant No. BK20181351, the Six Talent Peaks Project in Jiangsu Province under Grant No. JY-059, the Qinglan Project of Jiangsu Province of China, the National Natural Science Foundation of China under Grant No. 11975306, the Fundamental Research Fund for the Central Universities under the Grant Nos. 2019ZDPY07 and 2019QNA35, and the General Financial Grant from the China Postdoctoral Science Foundation under Grant Nos. 2015M570498 and 2017T100413.

\end{document}